

\magnification=\magstep1
\nopagenumbers \headline={\tenrm\hfil --\folio--\hfil}
\baselineskip=18pt  \lineskip=3pt minus 2pt   \lineskiplimit=1pt
\hsize=15true cm  \vsize=24true cm
\def\tomb{\phantom{.}\hfill\vrule height.4true cm width.3true cm
\par\smallskip\noindent}
\def\ker{{\rm Ker\,}}  \def\chop{\hfill\break}  
\def\f #1,#2.{\mathsurround=0pt \hbox{${#1\over #2}$}\mathsurround=5pt}
\def\s #1.{_{\smash{\lower2pt\hbox{\mathsurround=0pt $\scriptstyle
#1$}}\mathsurround=5pt}}
\def\r{{\hbox{\mathsurround=0pt$\rm I\! R$\mathsurround=5pt}}}
\def\mapdownl #1;{\vcenter{\hbox{$\scriptstyle#1$}}\Big\downarrow}
\def\mapdownr #1;{\Big\downarrow\rlap{$\vcenter{\hbox{$\scriptstyle#1$}}$}}
\tolerance=1600 \mathsurround=5pt  \def\aut{{\rm Aut}\,}
\def\maprightu #1;{\smash{\mathop{\longrightarrow}\limits^{#1}}}
\def\maprightd #1;{\smash{\mathop{\longrightarrow}\limits_{#1}}}

\def\brc #1,#2.{\left\langle #1\,|\,#2\right\rangle}
\def\rn#1{{\romannumeral#1}}  \def\cl #1.{{\cal #1}}
\def\convl #1,#2.{\mathrel{\mathop{\longrightarrow}\limits^{#1}_{#2}}}
\def\convr #1,#2.{\mathrel{\mathop{\longleftarrow}\limits^{#1}_{#2}}}
\def\set #1,#2.{\left\{\,#1\;\bigm|\;#2\,\right\}}
\def\theorem#1@#2@#3\par{\smallskip\parindent=.6true in \itemitem{\bf #1}
{\sl #2}\parindent=20pt\smallskip\itemitem{\it Proof:\/}#3\tomb}
\def\thrm#1"#2"#3\par{\smallskip\parindent=.6true in \itemitem{\bf #1}
{\sl #2}\parindent=20pt\smallskip\itemitem{\it Proof:\/}#3\tomb}
\def\teorem#1@#2@ {\smallskip\parindent=.6true in \itemitem{\bf Theorem #1}
{\sl #2}\hfill \parindent=20pt\smallskip\noindent} \def\ab{\allowbreak}
\def\fteorem#1@#2@ {\smallskip\parindent=.6true in \itemitem{\bf #1}{\sl #2}
\hfill\parindent=20pt\smallskip\noindent}    
     
\def\shift #1;{\mathord{\phantom{#1}}}  
\def\ref #1.{\mathsurround=0pt${}^{#1}\phantom{|}$\mathsurround=5pt}
\def\cross #1.{\mathrel{\jackup 3,\mathop\times\limits_{#1}.}}
\def\jackup#1,#2.{\raise#1pt\hbox{\mathsurround=0pt $#2$\mathsurround=5pt}}
\def\hlf{{\f 1,2.}}  \def\wt{\widetilde}

  \def\and{\wedge}
\def\bbrc #1,#2,#3.{\langle #1 |\,#2\,|#3\rangle}
   
\def\cst#1,{{C^*(#1-\un)}}
\def\un{\hbox{\mathsurround=0pt${\rm 1}\!\!{\rm 1}$\mathsurround=5pt}}

\def\alg#1.{{C^*(#1)}}

\def\ccr #1,#2.{{\overline{\Delta(#1,\,#2)}}}
\pageno=1  \noindent
\centerline{\bf Integrated Differential Geometry. Commutative and
Noncommutative.}
\vglue .3in
\centerline{Hendrik Grundling,}
\centerline{ Department of Pure Mathematics, University of
New South Wales,}
\centerline{ P.O. Box 1, Kensington, NSW 2033, Australia.}
\centerline{ email: hendrik@solution.maths.unsw.edu.au}
\vglue .2in
\itemitem{{\bf Abstract}}{\sl
For a manifold $M$ we define a structure on the group action
of ${\rm Diff}(M)$ on $C^\infty(M)$ which reduces to
the usual differential geometry upon differention at zero
along the one--parameter groups of ${\rm Diff}(M)$.
This ``integrated differential geometry'' generalises to all
group actions on associative algebras, including noncommutative
ones, and defines an
``integrated de Rham cohomology,'' which provides a new
set of invariants for group actions. We calculate the first few
integrated de Rham cohomologies for two examples:-
a discrete group action on a commutative algebra, and
a continuous Lie group action on a noncommutative matrix algebra. }
\chop
\par\noindent
{\bf Keywords:} group action, cohomology, noncommutative differential geometry.
\chop {\bf AMS classification: 46L87, 58B30, 46L55, 18G60, 14F99}
\chop{\bf Running headline:} Integrated Differential Geometry.
\vfill\eject

\beginsection 1. Introduction.

A problem with blending C*--algebras and differential geometry
as in the various approaches to noncommutative differential
geometry [1,2,3,4,5], is that C*--algebras
deal best with bounded information, whilst differential
geometry contains unbounded information expressed infinitesimally.
Connected to this is the fact that C*--algebras are appropriate to
the category of continuous function spaces with homeomorphisms,
 whilst differential
geometry is appropriate to smooth function spaces with diffeomorphisms.
This suggests that one should look for a larger ``integrated''
structure on a manifold, definable on its continuous functions,
 which can be ``differentiated at zero'' on the smooth functions
to reproduce the usual differential geometry associated with
the manifold. This larger structure can then be generalised to
noncommutative C*--algebras with greater ease,
thus avoiding derivations of dense *--algebras [4,6].

This paper runs as follows. In Sect. 2 we set up integrated differential
forms on a manifold, and generalise this to noncommutative algebras
in Sect. 3. On the set of these, we define an ``integrated''
differential $\hat d$ in Sect. 4, and show that it satisfies
$\hat d^2=0$ and reduces to the usual differential in the
case of an algebra of smooth functions on a manifold,
when we differentiate at zero on one--parameter subgroups of
${\rm Diff}(M)$. This defines then an ``integrated de Rham cohomology''
of which we calculate the first two for an example in Sect. 5
consisting of the shift automorphism acting on an algebra of sequences.
In Sect. 6 we work out the first ``integrated de Rham cohomology''
for the algebra $M_2({\bf C})$ under the action of the identity component
of its automorphism group.

\beginsection 2. The Basic Set--Up. Commutative Case.

Let $M$ be an $n\hbox{--dimensional}$ manifold, not necessarily compact.
Now the vector fields $X\in\cl X.(\cl B.)$ (i.e. the derivations of
the smooth algebra $\cl B.:=C_0^\infty(M)\subset C_0(M)=:\cl A.$, subscript
$0$ indicates functions vanishing at infinity) need not be complete
(i.e. integrable).
However, differential forms are fully defined on the vector fields of
compact support $\cl X._c(\cl B.)$, and the latter are indeed complete,
and form a Lie ideal of $\cl X.(\cl B.)$ which is a
$\cl B.\hbox{--module}$.
For $X\in\cl X._c(\cl B.)$, denote its flow by $\varphi^X:\r\to{\rm Diff}\,M$
which in turn defines a one--parameter automorphism group for the
C*--algebra $\cl A.=C_0(M)$ by
$$\left(\alpha^X_t(f)\right)(m):=f(\varphi^X_t(m))\qquad\forall\;
f\in C_0(M),\; t\in\r,\; m\in M$$
which clearly preserves $\cl B.$.

Consider the one--forms of $M$, but instead of using the usual definition
of $\cl B.\hbox{--linear}$ maps from $\cl X._c(\cl B.)$ to $\cl B.$,
we use the fact that any smooth one--form $\omega\in\Omega^1(M)$ has a
(nonunique) expression as $\omega=\sum\limits_ig_i\, df_i$; $g_i,\;\ab
f_i\in\cl B.$. Now for all $X\in\cl X._c(\cl B.)$, $m\in M$:
$$df(X)(m)=X(f)(m)={d\over dt}\, f(\varphi^X_t(m))\big|_{t=0}$$
and so for $\omega$:
$$\eqalignno{\omega(X)(m)&=\Big(\sum_ig_i\, df_i\Big)(X)(m)=
{d\over dt}\sum_ig_i(m)\, f_i(\varphi^X_t(m))\Big|_{t=0}\cr
&={d\over dt}\sum_ig_i(m)\,\alpha^X_t(f_i)(m)\Big|_{t=0}&(2.1)\cr}$$
This suggests the following:
\itemitem{\bf Definition:} An {\it integrated one--form}
$\wt\omega$ of $\omega\in\Omega^1(M)$ is a map
$\wt\omega:\r\times\cl X._c(\cl B.)\to\cl B.$ such that
$$\tau(\wt\omega)(X)(m):=
{d\over dt}\wt\omega(t,\, X)(m)\big|_{t=0}
=\omega(X)(m)\qquad\forall\,X\in\cl X._c,
\; m\in M$$

\noindent There may be several integrated one--forms for each one--form.
In particular, if $\omega=\sum\limits_ig_i\, df_i$ is a representation
of $\omega$, then by (2.1), $\wt\omega(t,\, X)=\sum\limits_ig_i\alpha_t^X(f_i)$
will be an integrated one--form for $\omega$.
We now limit our attention to those integrated one--forms coming
from representations $\omega=\sum g_i\,df_i$, as above:
\itemitem{\bf Definition:} Given the set of one--parameter groups
$\alpha:\r\times\cl X._c(\cl B.)\to\aut\cl A.$, define $\wt\Omega^1_\alpha$
as the set of all those integrated one--forms of the form
$$\wt\omega(t,\, X)=\sum_{i=1}^kg_i\alpha^X_t(f_i)\,,\quad f_i,\,g_i
\in\cl B.,\;k<\infty\,.$$

\noindent Note that $\wt\Omega_\alpha^1$ is a $\cl B.\hbox{--linear}$
space, and $\tau:\wt\Omega^1_\alpha\to\Omega(M)$ is a surjective
$\cl B.\hbox{--linear}$ map, i.e. $\tau(f\wt\omega)=f\,\tau(\wt\omega)$
for all $f\in\cl B.$, but since $\omega=\tau(\wt\omega)$ is
$\cl B.\hbox{--linear}$, we also have that
$$\tau(\wt\omega)(fX)(m)=f(m)\cdot\tau(\wt\omega)(X)(m)\qquad\forall\,
X\in\cl X._c(\cl B.),\;f\in\cl B.\;,$$
and this expresses locality w.r.t. $M$ (which may be lost in the
noncommutative case, cf. [5, 1]).
First, let us generalise away from the smooth structures on $M$
to the merely continuous:
\itemitem{\bf Definition:} Given any set of one--parameter groups
$\beta:\r\times I\to\aut\cl A.$ ($I$ is an index set), the set
$\hat\Omega^1_\beta$ of {\it total one--forms} of $\beta$
consists of all maps $\wt\omega:\r\times I\to\cl A.$ of the
form
$$\wt\omega(t,\, X)=\sum_ig_i\,\beta_t^X(f_i)\;,\qquad
f_i,\;g_i\in\cl A.\,;\;X\in I,\;t\in\r\,.$$
(Notation: $\beta_t^X:=\beta(t,\, X)$).
\item{\bf Notes: (1)} In the case $\alpha:\r\times\cl X._c(\cl B.)
\to\cl B.$ above, $\hat\Omega_\alpha^1\supset\wt\Omega^1_\alpha$,
where the extra elements come from choices $g_i,\;f_i\in
\cl A.\backslash\cl B.$. Clearly the map $\tau$ will only
be definable for those $\wt\omega$ where $\alpha_t^X(f_i)$
is differentiable in $t$ at zero for all $X\in\cl X._c(\cl B.)$
and $i$. Denote the set of these by $D^1(\tau)$.
\item{\bf (2)} $\hat\Omega^1_\beta$ is an $\cl A.\hbox{--module}$,
and depends on the choice of $\beta$.
\item{\bf (3)} Since $\tau:D^1(\tau)\to\Omega^1(M)$
maps integrated one--forms of the type $\wt\omega(t,\, X)=\alpha^X_t(f)$
to exact one--forms, we will later want to identify the exact
integrated one--forms as those of this type.

\noindent Next consider smooth two--forms. These also have (nonunique)
expressions $\omega=\sum\limits_ig_i\,df_i\and dh_i$ where $g_i,\;f_i,\;
h_i\in\cl B.$. Then
$$\eqalignno{\omega(X,\, Y)(m)&=\sum_ig_i(m)\,(df_i\and dh_i)(X,\, Y)(m)\cr
&=\hlf\sum_ig_i(m)\big[df_i(X)\cdot dh_i(Y)-dh_i(X)\cdot df_i(Y)\big](m)\cr
&={d\over dt}\cdot{d\over ds}\left\{\hlf\sum_ig_i(m)\,\big[
\alpha^X_t(f_i)\cdot\alpha_s^Y(h_i)-\alpha_s^Y(f_i)\cdot
\alpha_t^X(h_i)\big](m)\right\}\bigg|_{t=0=s}\cr}$$
which suggests that an integrated two--form $\wt\omega:(\r\times\cl X._c(\cl
B.)
)^2\to\cl B.$ for $\omega$ is
$$\wt\omega(t,X;\,s,Y)=\sum_ig_i\left(\alpha^X_t(f_i)\,\alpha_s^Y(h_i)-
\alpha_s^Y(f_i)\,\alpha_t^X(h_i)\right)\;.$$
In general the full set of integrated k--forms are:
\itemitem{\bf Definition 2.2:} Given the set of one--parameter groups
$\alpha:\r\times\cl X._c(\cl B.)\to\aut\cl A.$, the set $\wt\Omega^k_\alpha$
of {\it integrated k--forms} consists of all antisymmetric maps\chop
$\wt\omega:(\r\times\cl X._c(\cl B.))^k\to\cl B.$ such that
$${d\over dt_1}\cdots{d\over dt_k}\,\wt\omega(t_1,X_1;\ldots;\,t_k,X_k)
\bigg|_{t_i=0\;\forall i}=\omega(X_1,\ldots,\, X_k)(m)$$
defines a k--form $\omega\in\Omega^k(M)$.
\item{\bf Notes:} (1) Clearly for a k--form
$\omega=\sum\limits_ig_i\,df_i^1\and\cdots\and df_i^k$ one integrated k--form
is
$$\wt\omega(t_1,X_1;\cdots;\, t_k,X_k)=\sum_ig_i\sum_{\sigma\in{\bf S}_k}
\epsilon^\sigma\alpha_{t_{\sigma(1)}}^{X_{\sigma(1)}}(f_i^1)\cdots
\alpha^{X_{\sigma(k)}}_{t_{\sigma(k)}}(f_i^k)\;$$
where ${\bf S}_k$ is the permutation group of $k$ objects
and $\epsilon^\sigma$ is the parity of $\sigma\in{\bf S}_k$.
\item{\bf (2)} $\wt\Omega^k_\alpha$ is a $\cl B.\hbox{--linear}$ space and
we have as above, the surjective $\cl B.\hbox{--linear}$ map
$\tau:D^k(\tau)\to\Omega^k(M)$ given by
$$\eqalignno{\tau(\wt\omega)(X_1,\ldots,\, X_k)(m):=
{d\over dt_1}\cdots&{d\over dt_k}\,\wt\omega(t_1,X_1;\ldots;\,
t_k,X_k)\bigg|_{t_i=0\forall i}&\hbox{hence}\cr
\tau(\wt\omega)(X_1,\ldots,\,fX_i,\ldots,\,X_k)(m)&=
f(m)\cdot\tau(\wt\omega)(X_1,\ldots,\,X_k)(m)\quad\forall\,f\in
\cl B.,\;i\,,\cr}$$
where $D^k(\tau)\subset\wt\Omega^k_\alpha$ is the domain of $\tau$.

\noindent
Next we generalise the previous definitions away from both the
smooth structures, and from the one--parameter groups (hence derivations):
\itemitem{\bf Definition:} Given an indexed set of group actions
$\beta:G\times I\to\aut\cl A.$ ($I$ is an index set), the set
$\hat\Omega^k_\beta$ of {\it total k--forms} consists of all
maps\chop $\wt\omega:(G\times I)^k\to\cl A.$ of the form:
$$\wt\omega(g_1,X_1;\cdots;\,g_k,X_k)=
\sum_ih_i\sum_{\sigma\in{\bf S}_k}\epsilon^\sigma\beta^{X_{\sigma(1)}}\s
g_{\sigma(1)}.(f_i^1)\cdots\beta^{X_{\sigma(k)}}\s g_{\sigma(k)}.
(f_i^k)\eqno{-(*)}$$
for $h_i,\;f_i^\ell\in\cl A.$, $g_i\in G$, $X_i\in I$, and the sum over $i$
is finite.

\item{\bf Notes} (1) $\hat\Omega_\beta^k$ is
a left and right $\cl A.\hbox{--module}$, hence an $\cl A.\hbox{--linear}$
space, and depends on
$\beta$. An $\wt\omega\in\hat\Omega^k_\beta$ is specified as a
{\bf map} $\wt\omega:(G\times I)^k\to\cl A.$, so may have
different representations in the form $(*)$.
\item{(2)} Clearly there are canonical choices for $\beta$, e.g.
$G=\aut\cl A.$ and $I=\{1\}$. If there is more than one non--diffeomorphic
differential structure for $M$, there are two
dense $C^\infty\hbox{--subalgebras}$ $\cl B._1$ and $\cl B._2$ of
$\cl A.$ such that $\cl B._1$ and $\cl B._2$ are not in the same
automorphism class, then $G$ can be the diffeomorphism group with
respect to either of these.
\item{(3)} We lose locality information by allowing any group $G$.

\beginsection 3. Noncommutative integrated differential forms.

Maintain the concepts and notation of the last section.
In this section we would like to generalise the total k--forms
of the last section to noncommutative algebras $\cl A.$.
We remark that in the literature noncommutative differential
forms have already appeared, cf. [1,2], but here we
follow a different route.
We first examine the algebraic context of the
integrated differential forms. Following the line of thought above, observe
that a reasonable ``integrated covariant k--tensor'' will be a map
$\varphi:(G\times I)^k\to\cl A.$ of the form:
$$\varphi(g_1,X_1;\ldots;\,g_k,X_k)=\sum_ih_i\,
\beta^{X_1}\s g_1.(f_i^1)\cdots\beta^{X_k}\s g_k.(f_i^k)$$
and such maps form an algebra $\cl T.^k(\cl A.)$ under pointwise
multiplication:
$$(\varphi\cdot\psi)(g_1,X_1;\cdots;\, g_k,X_k)=
\varphi(g_1,X_1;\cdots;\,g_k,X_k)\cdot\psi(g_1,X_1;\cdots;\,g_k,X_k).$$
We also have the usual ${\bf N}\hbox{--graded}$ product $\star$ given by:
$$(\varphi\star\psi)(g_1,X_1;\ldots;\, g_m,X_m)=
\varphi(g_1,X_1;\ldots;\,g_k,X_k)\cdot\psi(g_{k+1},X_{k+1};\ldots;\,
g_m,X_m)$$
where $\varphi$ is a k--tensor and $\psi$ an $(m-k)\hbox{--tensor}$. However,
we will not need the $\star\hbox{--product}$ much.
Note that inside $\cl T.^k(\cl A.)$ the symmetric tensors is a
subalgebra whilst the antisymmetric tensors (the k--forms) is
a subspace. Now for the noncommutative generalisation, we henceforth
assume $\cl A.$ to be any associative *--algebra.
\itemitem{\bf Definition:} Given the *--algebra $\cl A.$ and
a fixed subgroup $G\subset\aut\cl A.$, let $\cl M.^k(G,\,\cl A.)$
be the space of maps $\varphi:G^k\to\cl A.$ and make it into a
*--algebra with pointwise operations. For a fixed $A\in\cl A.$,
define the elements $\varphi_0^A$ and $\varphi^A_\ell\in
\cl M.^k(G,\,\cl A.)$ by
$$\varphi_0^A(\alpha_1,\ldots,\,\alpha_k):=A\;,\qquad\quad
\varphi_\ell^A(\alpha_1,\ldots,\,\alpha_k):=\alpha_\ell(A)\,,$$
for $\alpha_i\in G$, $\ell\in\{1,\ldots,k\}$.
Then we define $\cl T.^k(\cl A.)$ as the *--algebra generated
in $\cl M.^k(G,\,\cl A.)$ by the set
$\set\varphi_\ell^A,{A\in\cl A.\,;\;\;\ell=0,\ldots,k}.$.
\item{\bf Notes}(1) We think of the elements of $\cl T.^k(\cl A.)$
as maps $\varphi:G^k\to\cl A.$ of the form
$$\varphi(\alpha_1,\ldots,\,\alpha_k)=\sum_iA_i^0\prod^{L_i}_{j=1}
\alpha\s s_i(j).(B^j_i)\,A_i^j\eqno(3.1)$$
for all $\alpha_\ell\in G$, where $A_i^\ell,\; B_i^\ell,\;
C_i^\ell\in\wt\cl A.$ are fixed and $s_i:\{1,2,\ldots,L_i\}\to
\{1,2,\ldots,k\}$ is a surjection, $L_i\geq k$.
$\wt\cl A.$ is $\cl A.$ if it has an identity, and it is
$\cl A.$ with the identity adjoined otherwise. So $\cl T.^k(\cl A.)$
excludes maps of the form $\varphi(\alpha_1,\,\alpha_2)=\alpha_1\circ
\alpha_2(A)$. Clearly $\cl T.^0(\cl A.)=\cl A.$.
Note that the same tensor $\varphi\in\cl T.^k(\cl A.)$ may
have more than one representation (3.1), given that it is defined as
a map.
\item{(2)} To recover the tensor algebra $\cl T.^k(\cl A.)$ of
above, for $\cl A.$ the continuous functions on a manifold,
we let $G={\rm Diff}\,M\subset\aut\cl A.$, and identify
the one--parameter groups of the compactly supported vector fields in
$G$.
\item{(3)} Note that $\cl T.^{k-r}(\cl A.)\subset\cl T.^k(\cl A.)$
for $0\leq r\leq k$ where a $\varphi\in\cl T.^{k-r}(\cl A.)$
is realised as a $k\hbox{--tensor}$ $\wt\varphi\in\cl T.^k(\cl A.)$
which is constant in the last $r$ variables, i.e.
$$\wt\varphi(\alpha_1,\ldots,\,\alpha_k)=\varphi(\alpha_1,
\ldots,\,\alpha_{k-r})\;.$$

\noindent On $\cl T.^k(\cl A.)$, identify the symmetrising and
antisymmetrising projectors:
$$\eqalignno{(P_+\varphi)(\alpha_1,\ldots,\,\alpha_k)&:=
\f 1,k!.\sum_{\sigma\in{\bf S}_k}\varphi(\alpha\s\sigma(1).,
\ldots,\,\alpha\s\sigma(k).)\cr
\Big(=\f 1,k!.\sum_{\sigma\in{\bf S}_k}\sum_iA_i^0&\prod_{j=1}^{L_i}
\alpha\s\sigma(s_i(j)).(B_i^j)\,A_i^j\qquad\hbox{for $\varphi$ as in (3.1)}
\;\Big)\cr
(P_-\varphi)(\alpha_1,\ldots,\,\alpha_k)&:=
\f 1,k!.\sum_{\sigma\in{\bf S}_k}\epsilon^\sigma\,\varphi(
\alpha\s\sigma(1).,\ldots,\,\alpha\s\sigma(k).)\cr}$$
for all $\varphi\in\cl T.^k(\cl A.)$ and $\alpha_i\in G$.
Obviously $(P_\pm)^2=P_\pm$. Define the symmetric (resp. antisymmetric)
tensors over $\cl A.$ by $\cl T.^k_\pm(\cl A.):=P_\pm\cl T.^k(\cl A.)$,
then  we regard $\hat\Omega^k:=\cl T._-^k(\cl A.)$ as the
{\it integrated k--forms over $\cl A.$} with respect to
$G\subset\aut\cl A.$. Note that $\hat\Omega^0=\cl A.$.
\item{\bf Notes}(1) Under pointwise multiplication
$\cl T.^k_+(\cl A.)\cdot\cl T._+^k(\cl A.)\subseteq\cl T._+^k(\cl A.)
\supseteq\cl T._-^k(\cl A.)\cdot\cl T._-^k(\cl A.)$ and
$\cl T._+^k(\cl A.)\cdot\cl T._-^k(\cl A.)\subseteq\cl T._-^k(\cl A.)
\supseteq\cl T._-^k(\cl A.)\cdot\cl T._+^k(\cl A.)$, and so
$\cl T.^k_+(\cl A.)\cup\cl T._-^k(\cl A.)$ generates a
${\bf Z}_2\hbox{--graded}$ *--algebra in $\cl T.^k(\cl A.)$.
\item{(2)} When $\cl A.=C_0(M)$, we regain the total k--forms
of the last section by replacing in the expression
$$\omega(\alpha_1,\ldots,\,\alpha_k)=\f 1,k!.\sum_{\sigma\in{\bf S}_k}
\epsilon^\sigma\sum_iA_i^0\prod^{L_i}_{j=1}\alpha\s\sigma(s_i(j)).
(B_i^j)\,A^j_i$$
$\alpha_i=\alpha\s t_i.^{X_i}$, $\f 1,k!.A_i^0\prod\limits^{L_i}_{j=1}
A_i^j=g_i$ and $\prod\limits_{j\in s^{-1}_i(\ell)}B^j_i=f_i^\ell$,
using commutativity in $\cl A.$.
\item{(3)} When $G=\aut\cl A.$, on a choice of one--parameter
subgroups $\alpha:\r\times I\to\cl A.$, we can define a map
$\tau$ to the infinitesimal k--forms as before (selecting a
domain in $\hat\Omega^k$), but we need to specify in what topology
the limits of the differentials should be taken. Possible
choices  are the
 C*--topology, weak operator topology of some representation of $\cl A.$,
weak *-topology w.r.t. some set of states etc. Note that by the definition
of the integrated k-forms, as maps from $G^k$ to $\cl A.$, there will be
some automatic continuity inherited from continuity of the action
of $G$ on $\cl A.$. In the case where we have a C*--dynamical system
in which the group is locally compact, this will be useful.

\beginsection 4. The integrated noncommutative de Rham complex.

In this section we want to define a de Rham structure
 on the integrated k--forms
$\hat\Omega^k$ of $\cl A.$ with respect to $G\subseteq\aut\cl A.$.
That is, we want a linear map\chop
$\hat d:\hat\Omega^k\to\hat\Omega^{k+1}$ such that $\hat d^2=0$,
making $\hat\Omega^*$ into a differential complex.
We want furthermore for the commutative case $\cl A.=C_0^\infty(M)$
that $\tau\circ\hat d$ be the usual exterior derivative for
differential forms. We will not expect $\hat d$ to be a derivation
with respect to the $\cl A.\hbox{--action}$ on $\hat\Omega^*$,
to enforce that
is the work of $\tau$ when it exists.

Consider the case when $\cl A.=C_0^\infty(M)$. Then a k--form
$\omega=\sum\limits_ig_i\,df_i^1\wedge\cdots\wedge df_i^k$
has differential $d\omega=\sum\limits_idg_i\wedge df^1_i\wedge
\cdots\wedge df_i^k$.
So if we take as in definition 2.2 an $\wt\omega\in\wt\Omega^k_\alpha$
given by
$$\wt\omega(t_1,X_i;\ldots;\,t_k,X_k)=\sum_ig_i
\sum_{\sigma\in{\bf S}_k}\epsilon^\sigma\,\alpha^{X_{\sigma(1)}}\s
t_{\sigma(1)}.(f_i^1)\cdots\alpha^{X_{\sigma(k)}}\s t_{\sigma(k)}.
(f_i^k)$$
then it seems reasonable to define $\wt d\,\wt\omega$ by analogy
$$\eqalignno{(\wt d\,\wt\omega)(t_1,X_1;\ldots;t_{k+1},X_{k+1})&:=
\f 1,k+1.\sum_{i=1}^\ell\sum_{\sigma\in{\bf S}_{k+1}}\epsilon^\sigma
\alpha^{X_{\sigma(1)}}\s t_{\sigma(1)}.(g_i)\cdot
\alpha^{X_{\sigma(2)}}\s t_{\sigma(2)}.(f^1_i)\cdots
\alpha^{X_{\sigma(k+1)}}\s t_{\sigma(k+1)}.(f^k_i)\cr}$$
and then we have that $\tau\circ\wt d=d$, the usual derivative.
However, a quick calculation with small $k$ shows that
$\wt d^2\not=0$. This will be fixed below, but we first
want to generalise $\wt d$ to $\cl T.^k(\cl A.)$ where
$\cl A.$ is noncommutative, and also ensure that $\wt d$ is
well--defined. Assuming now that $\cl A.$ is a general *--algebra,
let $\varphi\in\cl T.^k(\cl A.)$ have the representation
$$\varphi(\alpha_1,\ldots,\,\alpha_k)=\sum_iA_i^0\prod^{L_i}_{j=1}
\alpha\s s_i(j).(B_i^j)\,A_i^j$$
where $G\subset\aut\cl A.$ is fixed and $\alpha_\ell\in G$. Define:
$$(\wt d\varphi)(\alpha_1,\ldots,\,\alpha_{k+1}):=
\sum_i\alpha_{k+1}(A^0_i)\prod_{j=1}^{L_i}\alpha\s s_i(j).(B_i^j)\,
\alpha_{k+1}(A_i^j)\,.$$
However, due to the possible nonuniqueness of the representation
above for the map $\varphi:G^k\to\cl A.$, it is not clear that
$\wt d$ is well--defined. We rewrite the last expression
to get a more intrinsic expression:
$$\eqalignno{(\wt d\varphi)(\alpha_1,\ldots,\,\alpha_{k+1})&=
\alpha_{k+1}\Big(\sum_iA_i^0\prod_{j=1}^{L_i}\alpha_{k+1}^{-1}
\circ\alpha\s s_i(j).(B_i^j)A_i^j\Big)\cr
&=\alpha_{k+1}\left(\varphi(\alpha_{k+1}^{-1}\alpha_1,\ldots
,\,\alpha_{k+1}^{-1}\alpha_k)\right)&(4.1)\cr}$$
and clearly (4.1) is independent of the representation chosen
for $\varphi$, so we henceforth choose (4.1) as the definition
of $\wt d:\cl T.^k(\cl A.)\to\cl T.^{k+1}(\cl A.)$, which is
obviously well--defined. Now $\wt d$ is linear and we define
the odd and even parts of it by
$\wt d_\pm:=P^{(k+1)}_\pm\wt dP^{(k)}_\pm$, which clearly map
$\wt d_\pm:\cl T.^k_\pm(\cl A.)\to\cl T.^{k+1}_\pm(\cl A.)$.
Explicitly, let $\psi=P_-\varphi\in\hat\Omega^k$, then
$$\eqalignno{(\wt d_-P_-\varphi)(\alpha_1,\ldots,\,\alpha_{k+1})&=
(P_-\wt dP_-\varphi)(\alpha_1,\ldots,\,\alpha_{k+1})\cr
&=\f 1,(k+1)!.\sum_{\sigma\in{\bf S}_{k+1}}\epsilon^\sigma
(\wt dP_-\varphi)(\alpha\s\sigma(1).,\ldots,\,\alpha\s\sigma(k+1).)\cr
=\f 1,(k+1)!.\sum_{\sigma\in{\bf S}_{k+1}}\epsilon^\sigma
\alpha\s\sigma(k+1).&\left((P_-\varphi)(\alpha^{-1}\s\sigma(k+1).
\alpha\s\sigma(1).,\ldots,\,\alpha^{-1}\s\sigma(k+1).\alpha\s\sigma(k).)
\right)&(4.2)\cr}$$
Now observe that for any $\varphi\in\cl T.^k(\cl A.)$ we have:
$$\eqalignno{(\wt d\,\wt d\,\varphi)(\alpha_1,\ldots,\,\alpha_{k+2})&=
\alpha_{k+2}\left[(\wt d\varphi)(\alpha^{-1}_{k+2}\alpha_1,\ldots,\,
\alpha^{-1}_{k+2}\alpha_{k+1})\right]\cr
=\alpha_{k+2}\big[\alpha^{-1}_{k+2}\alpha_{k+1}[\varphi&(\alpha^{-1}_{k+1}
\alpha_{k+2}\alpha^{-1}_{k+2}\alpha_1,\ldots,\,\alpha^{-1}_{k+1}
\alpha_{k+2}\alpha_{k+2}^{-1}\alpha_k)]\big]\cr
&=\alpha_{k+1}\left[\varphi(\alpha^{-1}_{k+1}\alpha_1,\ldots,\,\alpha^{-1}_{k+1}
\alpha_k)\right]\cr
&=(\wt d\varphi)(\alpha_1,\ldots,\,\alpha_{k+1})\;.\cr}$$
Thus it is independent of $\alpha_{k+2}$.
Now to evaluate $(\wt d_-)^2$, let $\omega\in\hat\Omega^k=
\cl T.^k_-(\cl A.)$, then using (4.2):
$$\eqalignno{(\wt d_-\wt d_-\omega)(\alpha_1,\ldots,\alpha_{k+2})&=
\f 1,(k+2)!.\sum_{\sigma\in{\bf S}_{k+2}}\epsilon^\sigma\alpha\s\sigma(k+2).
\big((\wt d_-\omega)(\alpha^{-1}\s\sigma(k+2).\alpha\s\sigma(1).,\ldots
,\,\alpha^{-1}\s\sigma(k+2).\alpha\s\sigma(k+1).)\big)\cr
=\f 1,(k+2)!(k+1)!.\sum_{\sigma\in{\bf S}_{k+2}}\epsilon^\sigma
\alpha\s\sigma(k+2).&\bigg[\sum_{\bar\sigma\in{\bf S}_{k+1}}
\epsilon^{\bar\sigma}\alpha^{-1}\s\sigma(k+2).
\alpha\s\bar\sigma\sigma(k+1).\Big[\omega(\alpha^{-1}\s\bar\sigma
\sigma(k+1).\alpha\s\bar\sigma\sigma(1).,\ldots\cr
&\qquad\qquad\qquad\ldots,\,\alpha^{-1}\s\bar\sigma\sigma(k+1).
\alpha\s\bar\sigma\sigma(k).)\Big]\bigg]\cr
=\f 1,(k+1)!(k+2)!.\sum_{\sigma\in{\bf S}_{k+2}}\epsilon^\sigma
\sum_{\bar\sigma\in{\bf S}_{k+1}}&\epsilon^{\bar\sigma}
\alpha\s\bar\sigma\sigma(k+1).\left[\omega\big(\alpha^{-1}\s
\bar\sigma\sigma(k+1).\alpha\s\bar\sigma\sigma(1).,\ldots
,\,\alpha^{-1}\s\bar\sigma\sigma(k+1).\alpha\s\bar\sigma\sigma(k).
\big)\right]\cr
=\f 1,(k+1)!(k+2)!.\sum_{\sigma\in{\bf S}_{k+2}}&
\sum_{\bar\sigma\in{\bf S}_{k+1}}\epsilon^{\sigma\bar\sigma}
(\wt d\omega)(\alpha\s\bar\sigma\sigma(1).,\ldots,\,
\alpha\s\bar\sigma\sigma(k+1).)&(4.3)\cr}$$
Note that (4.3) is a linear combination of terms, each dependent
on only $k+1$ of the $k+2$ variables $\alpha_1,\ldots,\,\ab\alpha_{k+2}$.
\itemitem{\bf Definition:} Define the $\cl A.\hbox{--linear}$ maps
$\Delta_\ell:\cl T.^k(\cl A.)\to\cl T.^k(\cl A.)$, $1\leq\ell\leq k$
by $$(\Delta_\ell\varphi)(\alpha_1,\ldots,\,\alpha_k):=
\varphi(\alpha_1,\ldots,\alpha_k)-\varphi(\alpha_1,\ldots,\alpha_{
\ell-1},\,e,\,\alpha_{\ell+1},\ldots,\,\alpha_k)$$
where $e\in G\subseteq\aut\cl A.$ is the identity automorphism. Then
$$\Delta^{(k)}:=\Delta_1\Delta_2\cdots\Delta_k\;.$$

\noindent Note that a tensor $\varphi\in\cl T.^k(\cl A.)$
is independent of the $\ell^{th}$ entry, $\alpha_\ell$,
iff\chop $\Delta_\ell\varphi=0$. If $\varphi$ is independent
of any one of its arguments, then $\Delta^{(k)}\varphi=0$.
When it is obvious what degree tensor we are dealing with,
we will omit the superscript on $\Delta$. Using the the linearity
of $\Delta$, we see from (4.3) that $\Delta(\wt d_-\wt d_-\omega)=0$
$\forall\,\omega\in\hat\Omega^k$. Note that for $\omega
\in\hat\Omega^k$ we have by antisymmetry:
$$(\Delta\omega)(\alpha_1,\ldots,\,\alpha_k)=\omega(\alpha_1,\ldots,\alpha_k)
-\omega(e,\,\alpha_2,\ldots,\alpha_k)-\cdots-\omega(\alpha_1,\ldots,
\alpha_{k-1},e)$$
\thrm Theorem 4.4."Assume the hypotheses and notation above. Then\chop
$(\rn1)$ $\Delta^2=\Delta$,\chop
$(\rn2)$ $P_-\Delta=\Delta P_-$,\chop
$(\rn3)$ define $\hat d:\hat\Omega^k\to\hat\Omega^{k+1}$ by
$\hat d:=\Delta\wt d_-\Delta$, then $\hat d^2=0$."
$(\rn1)$ By definition $(\Delta\psi)(\alpha_1,\ldots,\alpha_k)=
\psi(\alpha_1,\ldots,\,\alpha_k)+{}$terms in which some of the
$\alpha_i\hbox{'s}$ have been replaced by $e$. The latter terms
are in $\ker\Delta$, so clearly
$$(\Delta\Delta\psi)(\alpha_1,\ldots,\,\alpha_k)
=(\Delta\psi)(\alpha_1,\ldots,\,\alpha_k)\quad\forall\,\psi\in\cl T.^k
(\cl A.)\,.$$
$(\rn2)$ Let $\psi\in\cl T.^k(\cl A.)$, then
$$\eqalignno{(\Delta\psi)(\alpha_1,\ldots,\,\alpha_k)&=
\psi(\alpha_1,\ldots,\,\alpha_k)-[\psi(e,\,\alpha_2,\ldots,\alpha_k)
+\cdots+\psi(\alpha_1,\ldots,\alpha_{k-1},e)]\cr
&\qquad\qquad+\sum_{i<j}\psi(\alpha_1,\ldots,\alpha_{i-1},e,\alpha_{i+1},
\ldots,\alpha_{j-1},e,\alpha_{j+1},\ldots,\alpha_k)-\cdots\cr
&=:\Big(1-\sum^k_{\alpha_i\to e}+\sum^k_{\alpha_i,\,\alpha_j\to e
\atop i< j}-\cdots\Big)\psi(\alpha_1,\ldots,\alpha_k)\cr}$$
in self--evident notation. So
$$\eqalignno{(P_-\Delta\psi)(\alpha_1,\ldots,\,\alpha_k)&=
\f 1,k!.\sum_{\sigma\in{\bf
S}_k}\epsilon^\sigma(\Delta\psi)(\alpha\s\sigma(1).,
\ldots,\,\alpha\s\sigma(k).)\cr
=\f 1,k!.\sum_{\sigma\in{\bf S}_k}\epsilon^\sigma\Big(1-&
\sum^k_{\alpha_{\sigma(i)}\to e}+\sum^k_{\alpha_{\sigma(i)},\,
\alpha_{\sigma(j)}\to e\atop i< j}-\cdots\Big)
\psi(\alpha\s\sigma(1).,\ldots,\,\alpha\s\sigma(k).)\,.\cr}$$
Observe that for each $\sigma\in{\bf S}_k$ we have
$$\eqalignno{\sum^k_{\alpha_{\sigma(i)}\to e}\psi(\alpha\s\sigma(1).,
\ldots,\,\alpha\s\sigma(k).)&=\sum^k_{\alpha_i\to e}\psi(\alpha\s\sigma(1).,
\ldots,\,\alpha\s\sigma(k).)&\hbox{and}\cr
\sum^k_{\alpha_{\sigma(i)},\,\alpha_{\sigma(j)}\to e\atop
i<j}\psi(\alpha\s\sigma(1).,\ldots,\,\alpha\s\sigma(k).)&=
\sum^k_{\alpha_i,\,\alpha_j\to e\atop i<j}\psi(\alpha\s\sigma(1).,
\ldots,\,\alpha\s\sigma(k).)\cr}$$
since the sums are over all possible single replacements or pairs of
replacements by $e$. Similar statements are
true for the higher terms. Thus
$$\eqalignno{(P_-\Delta\psi)(\alpha_1,\ldots,\,\alpha_k)&=
\f 1,k!.\sum_{\sigma\in{\bf S}_k}\epsilon^\sigma\Big(1-
\sum^k_{\alpha_i\to e}+\sum^k_{\alpha_i,\,\alpha_j\to e\atop i<j}
-\cdots\Big)\psi(\alpha\s\sigma(1).,\ldots,\,\alpha\s\sigma(k).)\cr
&=\Big(1-\sum^k_{\alpha_i\to e}+\sum^k_{\alpha_i,\,\alpha_j\to e\atop
i<j}-\cdots\Big)\f 1,k!.\sum_{\sigma\in{\bf S}_k}\epsilon^\sigma
\psi(\alpha\s\sigma(1).,\ldots,\,\alpha\s\sigma(k).)\cr
&=(\Delta P_-\psi)(\alpha_1,\ldots,\,\alpha_k)\;.\cr}$$
$(\rn3)$ To show that $\hat d^2=\Delta\wt d_-\Delta\wt d_-\Delta=0$
on $\hat\Omega^k$, recall the result $\Delta\wt d_-\wt d_-=0$
from above. Then we will have that $\hat d^2=0$ if we can show
that $\Delta\wt d_-(1-\Delta)\wt d_-\Delta=0$ on $\hat\Omega^k$.
Now recall that $\wt d_-=P_-\wt dP_-$ maps into $\hat\Omega^{k+1}$
and that $1-\Delta$ on $\omega\in\hat\Omega^{k+1}$ has the form
$\sum\limits^{k+1}_{\alpha_i\to e}\omega(\alpha_1,\ldots,\,\alpha_{k+1})$,
so it produces a sum with terms, each depending on only $k$ of the
$k+1$ original variables. Thus, since $\wt d_-$ can only add one
dependence, we find that $\wt d_-(1-\Delta)\wt d_-\Delta$ is
a sum of terms, each depending on only $k+1$ of the $k+2$
variables. Hence $\Delta\wt d_-(1-\Delta)\wt d_-\Delta=0$, i.e.
$\hat d^2=0$. This can also be done by explicit calculation.

\item{\bf Remarks:}(1) By this theorem we have obtained a new chain
complex, hence cohomology theory which can be associated to any group action on
an associative algebra $\cl A.$. Since $\aut\cl A.$ is an intrinsic
group action for $\cl A.$, the cohomology of $\cl A.$ w.r.t the group
$\aut\cl A.$ is an invariant of $\cl A.$.
\item{(2)} In the case where $\cl A.=C_0(M)$, and $G=\aut\cl A.={\rm
Homeo}(M)$, we thus have obtained a cohomology theory for any
topological space $M$, independent of differential structures.
In particular, let $M$ be a space having more than one differential
structure, e.g. $\r^4$, so we have two dense $C^\infty$ subalgebras
$\cl B._1,\;\cl B._2\subset\cl A.$ corresponding to nondiffeomorphic
differential structures with automorphism groups ${\rm Diff}_1M$
and ${\rm Diff}_2M$ resp.
 Then the integrated de Rham cohomologies
for $M$
are simply the restrictions of the cohomology for $\cl A.$
with $\aut\cl A.$ to the algebras $\cl B._i$ with groups
${\rm Diff}_iM$, $i=1,\;2$ and under the corresponding
$\tau\hbox{--maps}$ these map to the de Rham comomologies.
So the integrated de Rham cohomology of $\cl A.$ with
$\aut\cl A.$ is some sort of ``universal receptacle''
for all the de Rham cohomologies associated with $M$.
\item{(3)} Observe that the range of $\Delta$ consists of those
$\omega\in\cl T.^k$ for which $\omega(\alpha_1,\ldots,\,\alpha_k)=0$
if any $\alpha_i$ is equal to $e$. That is:
$$\eqalignno{\Delta\cl T.^k(\cl A.)&=
\set\omega\in\cl T.^k(\cl A.),\ker\omega\supset
\{e\}\times G\times\cdots\times G\cup\cdots\cup G\times\cdots\times
G\times\{e\}.\cr}$$
\thrm Theorem 4.5." Let $M$ be a finite dimensional manifold,
$\cl A.=C_0^\infty(M)$, $G={\rm Diff}_0M$. Then\chop
$(\rn1)$ there is a linear map $\tau:\hat\Omega^k\to\Omega^k(M)$
for each $k$, such that $\tau\circ\hat d=d\circ\tau$, where $d$
is the usual exterior derivative.\chop
$(\rn2)$ Denote the (integrated de Rham) cohomology produced by
the differential complex ${(\hat\Omega^*,\,\hat d)}$ by
$\hat H^*(\cl A.)$, then there is a surjective homomorphism
to the usual de Rham cohomology groups
$\hat\tau:\hat H^k(\cl A.)\to H^k(M)$ for each $k$."
$(\rn1)$ Let $\omega\in\hat\Omega^k$ and $X_1,\ldots,\,X_k\in
\cl X._c(\cl A.)$, and define
$$(\tau\omega)(X_1,\ldots,\,X_k)(m):={d\over dt_1}\cdots
{d\over dt_k}\,\omega(\alpha_{t_1}^{X_1},\ldots,\,\alpha_{t_k}^{X_k})
(m)\bigg|_{t_i=0\,\forall i}$$
where $\alpha^X_t$ is as before. Clearly $\tau$ is linear.
Now $\hat d\omega=\Delta\wt d_-\Delta\omega=\wt d_-\omega+{}$terms depending
on fewer than $k$ variables. Then since $\tau$ is zero on the latter terms,
$$\eqalignno{\big(\tau(\hat d\omega)\big)(X_1,\ldots,\,X_{k+1})(m)&=
{d\over dt_1}\cdots{d\over dt_{k+1}}(\hat d\omega)\big(\alpha^{X_1}_{t_1},
\ldots,\,\alpha^{X_{k+1}}_{t_{k+1}}\big)\bigg|_{t_i=0\,\forall i}\cr
&=\tau(\wt d_-\omega)(X_1,\ldots,\,X_{k+1})(m)\cr
&=(d\omega)(X_1,\ldots,\,X_{k+1})(m)\cr}$$
where the last equality is already known. Thus $\tau\circ\hat d
=d\circ\tau$.\chop
$(\rn2)$ This part follows from $(\rn1)$, since clearly if
$\hat d\omega=0$ then $d(\tau\omega)=0$, hence $\tau$ maps
closed forms to closed forms, and by the explicit formulae
above for forms, we see that any closed de Rham form is
an image under $\tau$ of a closed total form. Moreover if $\tau\omega$ is
exact, i.e. $\tau\omega=d\varphi=d\tau\wt\varphi=\tau\cdot\hat d\wt\varphi$,
then $\tau\omega$ is the image under $\tau$ of an exact form in
$\hat\Omega^*$.
Thus $\tau$ respects cohomology classes,
 so $\tau$ lifts to a linear surjective map by
$\hat\tau[\omega]:=[\tau\omega]$ where $[\omega]$ is the class
of $\omega\in\hat Z^k$, as claimed.

\item{\bf Notes}(1) These two theorems establish the claim that the
current constructions produce a cohomology theory which
generalises de Rham cohomology. It is also clear that the cohomology
classes $\hat H^*(\cl A.)$ are nontrivial for some $\cl A.$,
or else $\hat\tau$ will not map onto de Rham cohomology.
\item{(2)} Observe that the kernel of $\tau$ contains
all forms which are invariant in some entry.
The invariant forms in $\hat\Omega^1$, as an algebra under
pointwise operations, is isomorphic to $\cl A.$.
For the forms in $\hat\Omega^k$, $k\geq2$, invariance in
one slot automatically implies that such forms vanish
by antisymmetry.
\item{(3)} This cohomology is relevant for {\it actions}, hence
it can be used to study a single automorphism by letting
$G\subset\aut\cl A.$ be the group generated by that
automorphism.
Thus it can be used to study operators on linear spaces and
transformations on topological spaces (with the choice $\cl A.=
C_b(X)$ or $C_0(X)$ if $X$ is locally compact)
 without recourse to measure theory.
\item{(4)} Once we have equipped $\hat\Omega^*$ with a wedge
product in the obvious way, we do not expect it to be a differential algebra
with respect to $\hat d$. That can only be expected at the
infinitesimal level, once we have a map $\tau$ as in theorem 4.5.
Some homomorphic property of $\hat d$  remains, cf. (6.0) below.
\item{(5)} Whilst in general it may be hard to compute the
integrated cohomology $\hat H^*$, in the case of a Lie group
action $G\subset\aut\cl A.$ it may be easier to calculate
the {\it infinitesimal} cohomology, for which we would need
a $\tau\hbox{--map}$:
$${d\over dt_i}\cdots{d\over dt_k}\,\omega(\alpha_{t_1}^{(1)}
,\ldots,
\alpha_{t_k}^{(k)})\Big|_{t_i=0\,\forall i}=:(\tau\omega)(X_i,\ldots,X_k)
$$
where $\alpha_t^{(n)} =\exp tX_n$ are one--parameter groups in $G$.
To make sense of this, some topology on $\cl A.$ must have been given.
The $\tau\hbox{--map}$ then produces the infinitesimal
(possibly noncommutative!) de Rham cohomology from $\hat H^*$.
This will be done in the second example below in Sect.6.

\beginsection 5. Examples: A Discrete Action on a Commutative
Algebra.

Next we wish to work out $\hat H^1$ and $\hat H^2$
for some concrete examples, but before doing so, first
need the general formulii for closed and exact forms.
\thrm Lemma 5.1."Let $\cl A.$ be an associative algebra, and
$G\subseteq\aut\cl A.$ given. Then
a form $\omega\in\hat\Omega^1$ is\chop
$(\rn1)$ exact iff $\omega(\alpha)=\alpha(A)-A$ for all $\alpha\in G$
and some fixed $A\in\cl A.$ (depending on $\omega$)\chop
$(\rn2)$ closed iff for all $\alpha,\,\alpha_0\in G$:
$$\eqalignno{\alpha_0(\omega(\alpha_0^{-1}\cdot\alpha))&
-\alpha_0(\omega(\alpha_0^{-1}))
+\omega(\alpha_0)=\alpha(\omega(\alpha^{-1}\cdot\alpha_0))-\alpha(
\omega(\alpha^{-1}))+\omega(\alpha)\;.\cr}$$  "
$(\rn2)$ We first prove the second part. Now $\omega\in\hat\Omega^1$
is closed if $(\hat d\omega)(\alpha,\,\alpha_0)=0$ for all $\alpha,\,
\alpha_0\in G$. Expand this equation:
$$\eqalignno{0=&(\hat d\,\omega)(\alpha,\,\alpha_0)=
(\Delta P_-\wt d\,P_-\Delta\omega)(\alpha,\,\alpha_0)\cr
&=(P_-\wt d\,P_-\Delta\omega)(\alpha,\alpha_0)-
(P_-\wt d\,P_-\Delta\omega)(e,\alpha_0)-
(P_-\wt d\,P_-\Delta\omega)(\alpha,\,e)\cr
&=\hlf\big[(\wt d\,P_-\Delta\omega)(\alpha,\alpha_0)-(\wt d\,P_-\Delta\omega)(
\alpha_0,\alpha)-(\wt d\,P_-\Delta\omega)(e,\alpha_0)+
(\wt d\,P_-\Delta\omega)(\alpha_0,e)\cr
&\qquad\qquad\qquad-(\wt d\,P_-\Delta\omega)(\alpha,\,e)
+(\wt d\,P_-\Delta\Omega)(e,\,\alpha)\big]\cr
&=\hlf\Big[\alpha_0\big((P_-\Delta\omega)(\alpha_0^{-1}\alpha)\big)
-\alpha\big((P_-\Delta\omega)(\alpha^{-1}\alpha_0)\big)
-\alpha_0\big((P_-\Delta\omega)(\alpha_0^{-1})\big)\cr
&\qquad\qquad+(P_-\Delta\omega)(\alpha_0)
-(P_-\Delta\omega)(\alpha)+\alpha\big((P_-\Delta\omega)(\alpha^{-1})
\big)\Big]\cr
&=\hlf\big[\alpha_0\big(\omega(\alpha_0^{-1}\cdot\alpha)-\omega(e)\big)
-\alpha\big(\omega(\alpha^{-1}\cdot\alpha_0)-\omega(e)\big)
-\alpha_0\big(\omega(\alpha_0^{-1})-\omega(e)\big)\cr
&\qquad+\omega(\alpha_0)-\omega(e)-\omega(\alpha)+\omega(e)
+\alpha(\omega(\alpha^{-1})-\omega(e))\big]\cr
&=\hlf\big[\alpha_0(\omega(\alpha_0^{-1}\alpha))-\alpha(\omega(\alpha^{-1}
\alpha_0))-\alpha_0(\omega(\alpha_0^{-1}))+\alpha(\omega(\alpha^{-1}))
+\omega(\alpha_0)-\omega(\alpha)\big]\cr}$$
which proves $(\rn2)$.\chop
$(\rn1)$ $\omega\in\hat\Omega^1$ is exact if $\omega=\hat d\varphi$
for some $\varphi\in\hat\Omega^0=\cl A.$. Say $\varphi=A$, then
$$\eqalignno{\omega(\alpha)&=(\hat d\varphi)(\alpha)=
(\Delta P_-\wt d\,P_-\Delta\varphi)(\alpha)\cr
&=(\wt d\,P_-\Delta\varphi)(\alpha)-(\wt d\,P_-\Delta\varphi)(e)
=\alpha(A)-A\cr}$$

\item{\bf Notes:}(1) Observe that if a one--form is invariant
(i.e. $\omega(\alpha)=\omega(e)$), then it is closed,
but obviously the only exact invariant one--form is zero. So the
algebra $\cl A.$ itself will always constitute part of
$\hat H^1(\cl A.)$, corresponding to the invariant one--forms.
Forms of the type $\omega(\alpha)=\sum\limits_i[B_i\,\alpha(B_i)
+\alpha(B_i)\,B_i]$, $B_i\in\cl A.$ are always closed, as we
can check from (5.1\rn2).
\item{(2)} A zero--form $\varphi=A\in\hat\Omega^0=\cl A.$
is closed iff $(\hat d\varphi)(\alpha)=\alpha(A)-A=0$,
i.e. $A$ is $G\hbox{--invariant}$. Hence $\hat H^0(\cl A.)=\cl A.^G$.
\thrm Lemma 5.2."Given an associative algebra $\cl A.$ and group
$G\subseteq\aut\cl A.$, a two--form $\omega\in\hat\Omega^2$ is\chop
$(\rn1)$ exact whenever there is a $\varphi\in\hat\Omega^1$
such that for all $\alpha_i\in G$:
$$\eqalignno{\omega(\alpha_1,\,\alpha_2)&
=\alpha_2\big(\varphi(\alpha_2^{-1}
\alpha_1)-\varphi(\alpha_2^{-1})\big)-\alpha_1\big(\varphi(
\alpha_1^{-1}\alpha_2)-\varphi(\alpha_1^{-1})\big)
+\varphi(\alpha_2)-\varphi(\alpha_1)\cr
&\quad(=2(\hat d\varphi)(\alpha_1,\,\alpha_2)\;\;)\cr}$$
$(\rn2)$ closed whenever for all $\alpha_i\in G$:
$$\eqalignno{0&=\omega(\alpha_3,\,\alpha_2)+\omega(\alpha_1,\,\alpha_3)
+\omega(\alpha_2,\,\alpha_1)+\alpha_1\big[\omega(\alpha_1^{-1}\alpha_2,
\,\alpha_1^{-1}\alpha_3)+\omega(\alpha_1^{-1}\alpha_3,\,\alpha_1^{-1})\cr
&\quad+\omega(\alpha_1^{-1},\,\alpha_1^{-1}\alpha_2)\big]
+\alpha_2\big[\omega(\alpha_2^{-1}\alpha_3,\,\alpha_2^{-1}\alpha_1)+
\omega(\alpha_2^{-1},\,\alpha_2^{-1}\alpha_3)+\omega(\alpha_2^{-1}
\alpha_1,\,\alpha_2^{-1})\big]\cr
&\qquad\qquad+\alpha_3\big[\omega(\alpha_3^{-1}\alpha_1,\,
\alpha_3^{-1}\alpha_2)+\omega(\alpha_3^{-1}\alpha_2,\,\alpha_3^{-1})
+\omega(\alpha_3^{-1},\,\alpha_3^{-1}\alpha_1)\big]\;.\cr}$$"
$(\rn1)$ Now $\omega$ is exact iff $\omega=\hat d\varphi$ for some
$\varphi\in\hat\Omega^1$. Expand this equation, using (4.4\rn2) and
the fact that $P_-$ on $\hat\Omega^k$ is just the identity:
$$\eqalignno{\omega(\alpha_1,\,\alpha_2)&=(\hat d\,\varphi)(\alpha_1,\,
\alpha_2)=(\Delta P_-\wt d\,P_-\Delta\varphi)(\alpha_1,\,\alpha_2)
=(\Delta P_-\wt d\,\Delta\varphi)(\alpha_1,\,\alpha_2)\cr
&=(P_-\wt d\,\Delta\varphi)(\alpha_1,\,\alpha_2)-(P_-\wt d\Delta
\varphi)(\alpha_1,\,e)-(P_-\wt d\Delta\varphi)(e,\,\alpha_2)\cr
&=\hlf\big[(\wt d\Delta\varphi)(\alpha_1,\,\alpha_2)-(\wt d\Delta\varphi)
(\alpha_2,\,\alpha_1)-(\wt d\Delta\varphi)(\alpha_1,\,e)+
(\wt d\Delta\varphi)(e,\,\alpha_1)\cr
&\qquad\qquad\quad-(\wt d\Delta\varphi)(e,\,\alpha_2)+
(\wt d\Delta\varphi)(\alpha_2,\,e)\big]\cr
&=\hlf\big[\alpha_2\big((\Delta\varphi)(\alpha_2^{-1}\alpha_1)\big)-
\alpha_1\big((\Delta\varphi)(\alpha^{-1}_1\alpha_2)\big)
-(\Delta\varphi)(\alpha_1)+\alpha_1\big((\Delta\varphi)(\alpha_1^{-1})\big)\cr
&\qquad\qquad\quad-\alpha_2\big((\Delta\varphi)(\alpha_2^{-1})\big)
+(\Delta\varphi)(\alpha_2)\big]\cr
&=\hlf\big[\alpha_2\big(\varphi(\alpha_2^{-1}\alpha_1)\big)
-\alpha_2(\varphi(e))-\alpha_1\big(\varphi(\alpha_1^{-1}\alpha_2)\big)
+\alpha_1(\varphi(e))-\varphi(\alpha_1)+\varphi(e)\cr
&\quad+\alpha_1\big(\varphi(\alpha_1^{-1})\big)-\alpha_1(\varphi(e))
-\alpha_2\big(\varphi(\alpha_2^{-1})\big)+\alpha_2(\varphi(e))
+\varphi(\alpha_2)-\varphi(e)\big]\cr
&=\hlf\big[\alpha_2\big(\varphi(\alpha_2^{-1}\alpha_1)-
\varphi(\alpha_2^{-1})\big)-\alpha_1\big(\varphi(\alpha_1^{-1}\alpha_2)
-\varphi(\alpha_1^{-1})\big)+\varphi(\alpha_2)-\varphi(\alpha_1)\big]\cr}$$
which proves $(\rn1)$.\chop
$(\rn2)$ This is proven by (a lengthy) expansion of
$$0=(\hat d\omega)(\alpha_1,\,\alpha_2,\,\alpha_3)=
(\Delta P_-\wt d\,\Delta\omega)(\alpha_1,\,\alpha_2,\,\alpha_3)$$
which we omit as straightforward algebra.

\item{\bf Notes:} Observe that $\omega\in\hat\Omega^2$ is closed if
it satisfies
$$\omega(\alpha_3,\,\alpha_2)+\omega(\alpha_1,\,\alpha_3)+\omega(
\alpha_2,\,\alpha_1)=0\;.\eqno(5.3)$$
Another kind of closed form (of all orders) can be deduced from
the representation (3.1) of forms, when the $A_i^\ell$ are all
$G\hbox{--invariant.}$ At the level of maps, this condition
will read for such an $\omega\in\hat\Omega^k$:
$$\alpha_1\big(\omega(\alpha_1^{-1}\alpha_2,\ldots,\,\alpha_1^{-1}
\alpha_{k+1})\big)=\omega(\alpha_2,\ldots,\,\alpha_{k+1})$$
for all $\alpha_i\in G$, in which case $\wt d\,\omega=0$
and so $\hat d\,\omega=0$. For the case in theorem 4.5, these
closed forms map under $\tau$ to the exact de Rham forms.

\noindent Now we are ready to do examples.
\chop {\bf Example 0:}\chop
Let $\beta:H\to\aut\cl A.$ be a trivial action on an associative
algebra $\cl A.$, i.e. $\beta(H)=\iota=G$. Then we have
$\hat\Omega^0=\cl A.=\hat\Omega^1$ and $\hat\Omega^k=0$ for all $k\geq 2$.
Thus $\hat H^0(\cl A.)=\cl A.=\hat H^1(\cl A.)$
 and $\hat H^k(\cl A.)=0$ for all
$k\geq 2$.\chop
\noindent{\bf Example 1:}\chop
Here we want to study a single homeomorphism $T:X\to X$
of a locally compact space $X$. Let $\cl A.=C_0(X)$ and
define the automorphism $\alpha(f)(x):=f(Tx)$ for all
$f\in\cl A.$. Let $G$ be the group generated by $\alpha$
in $\aut\cl A.$, which is obviously a factor group of
${\bf Z}$. Then a general one--form $\omega\in\hat\Omega^1$
has an expression
$$\omega(\alpha^n)(x)=\sum_{\ell=1}^L\left\{f^\ell(x)\,\alpha^n(
h^\ell)(x)+\alpha^n(g^\ell)(x)+k^\ell(x)\right\}$$
for all $n\in{\bf Z}$ and fixed $f^\ell,\,h^\ell,\,g^\ell,\,k^\ell
\in\cl A.$. Closed forms must satisfy 5.1\rn2, i.e.
for all $n,\,m\in{\bf Z}$:
$$\eqalignno{\alpha^n(\omega(\alpha^{m-n}))-\alpha^n(\omega(\alpha^{-n}))
+\omega(\alpha^n)=\alpha^m&(\omega(\alpha^{n-m}))-\alpha^m(\omega(\alpha^{-m}))
+\omega(\alpha^m)\qquad\hbox{so}\cr
\sum_\ell^L\left\{\alpha^n(f^\ell)\alpha^m(h^\ell)-\alpha^n(f^\ell)h^\ell
+f^\ell\alpha^n(h^\ell)\right\}&=\sum^L_\ell\left\{\alpha^m(f^\ell)\alpha^n
(h^\ell)-\alpha^m(f^\ell)h^\ell+f^\ell\alpha^m(h^\ell)\right\}\cr}$$
Exact one--forms are of the type $\omega(\alpha^n)=\alpha^n(f)-f$.\chop
In particular, let us work out the first and second cohomology classes
for the shift operator on ${\bf Z}$. That is, we set $X={\bf Z}$,
$T:{\bf Z}\to{\bf Z}$ by $Tn=n+1$, so $\cl A.=C_0({\bf Z})$ consists
of sequences $f=\{f_i\}_{i\in{\bf Z}}$ which go to zero at both ends,
and pointwise multiplication is $f\cdot g=\{f_i\}\cdot\{g_j\}
=\{f_ig_i\}_{i\in{\bf Z}}$ in $\cl A.$. Note that $\cl A.$ has no
nonzero elements invariant under $\alpha$, and a general one--form
$\omega$ has now an expression
$$\omega(\alpha^n)=\sum_{\ell=1}^L\left\{f_i^\ell\,h^\ell_{i-n}
+g^\ell_{i-n}+k^\ell_i\right\}_{i\in{\bf Z}}\qquad\forall\,n$$
and hence $\omega$ is closed iff
$$\sum_\ell^L\left\{f_{i-n}^\ell h^\ell_{i-m}-f^\ell_{i-n}h_i^\ell
+f_i^\ell h^\ell_{i-n}\right\}_{i\in{\bf Z}}=
\sum_\ell^L\left\{f^\ell_{i-m}h^\ell_{i-n}-f^\ell_{i-m}h^\ell_i
+f^\ell_ih^\ell_{i-m}\right\}_{i\in{\bf Z}}$$
and on equating each entry separately, we find:
$$\sum_\ell^L\left\{f^\ell_{i-n}(h^\ell_{i-m}-h^\ell_i)
+f^\ell_i(h^\ell_{i-n}-h^\ell_{i-m})+f^\ell_{i-m}
(h_i^\ell-h^\ell_{i-n})\right\}=0$$
for all $n,\,m,\,i\in{\bf Z}$. Taking the limit $n\to\infty$:
$$\leqalignno{\sum_\ell^L\left\{-f_i^\ell h^\ell_{i-m}+
f^\ell_{i-m}h^\ell_i\right\}&=0\qquad\forall\,i,\,m\in{\bf Z}\cr
\sum_\ell^Lf_i^\ell h_j^\ell=\sum_\ell^Lf^\ell_jh^\ell_i&
\qquad\forall\,i,\,j\in{\bf Z}&\hbox{i.e.}\cr}$$
Note that this condition is also sufficient for $\omega$ to be
closed. Thus the closed one--forms $\omega$, are precisely those
which can be written:
$$\omega(\alpha^n)=\sum_{\ell=0}^L\left\{f_i^\ell h^\ell_{i-n}
+f^\ell_{i-n}h_i^\ell+g^\ell_{i-n}+k_i^\ell\right\}_{i\in{\bf Z}}
\eqno{ (5.3)}$$
for all $n\in{\bf Z}$. Denote the space of these by $\hat Z^1$.
 The exact one--forms are of the type
$\omega(\alpha^n)={\left\{f_{i-n}-f_i\right\}_{i\in{\bf Z}}}$ and
the space of these is
$$\hat B^1=(\alpha^{\bf Z}-\iota)(\cl A.):=\set\alpha^n(A)-A,{
A\in\cl A.,\;n\in{\bf Z}}.\;.\eqno{(5.4)}$$
\thrm Claim 5.5." There is a linear bijection from $\hat H^1(\cl A.)$
to the linear spaces\chop
$(\rn1)$ $\cl L.\subset\hat Z^1$ consisting of those $\omega$ of the form
$$\omega(\alpha^n)=\Big\{\sum^L_{\ell=1}(f_i^\ell h^\ell_{i-n}
+f_{i-n}^\ell h_i^\ell)+k_i\Big\}_{i\in{\bf Z}}\qquad(5.6)\quad\hbox{and}
$$
$(\rn2)$ $L[(\hat B^1)\cdot(\hat B^1
)]+\cl A.\subset\hat Z^1$ consisting of
$\omega$ of the form
$$\omega(\alpha^n)=\Big\{\sum_{\ell=1}^L(f^\ell_{i-n}-f_i^\ell)
(h^\ell_{i-n}-h^\ell_i)+k_i\Big\}_{i\in{\bf Z}}\;,\eqno{(5.7)}$$
$\{f_i^\ell\},\;\{h_i^\ell\},\;\{k_i\}\in\cl A.$."
$(\rn1)$  From (5.3) and (5.4) we see that every $\omega\in\hat Z^1$
is cohomologous to an element of $\cl L.$, in fact $\cl L.$ is in
$\hat Z^1$ and has nonempty intersection with each cohomology
class in $\hat Z^1$. We only
need to show that an $\omega\in\cl L.$ is exact iff it is zero.
Let $\omega\in\cl L.\cap \hat B^1$, so there is a $\varphi=
\{g_i\}_{i\in{\bf Z}}$ such that $\omega=\hat d\varphi$
and $\omega$ is of the form (5.6), i.e.
$$\sum_{\ell=1}^L(f_i^\ell h^\ell_{i-n}+f^\ell_{i-n}h^\ell_i)
+k_i=g_{i-n}-g_i\eqno{(5.8)}$$
for all $i,\,n\in{\bf Z}$. Let $n\to\infty$ to find
$k_i=-g_i$, so on substitution into (5.8) and replacing $i$
with $i+n$:
$$\sum^L_\ell(f^\ell_{i+n}h_i^\ell+f_i^\ell h^\ell_{i+n})=g_i
\qquad\forall\,i,\,n$$
so that in the limit $n\to\infty$ we find $g_i=0=k_i$ for all $i$,
hence $\omega(\alpha^n)=\hat d\{g_i\}=\{g_{i-n}-g_i\}_{i\in
{\bf Z}}=0$.\chop
$(\rn2)$ Now every element of $\cl L.$ is cohomologous to
an element of\chop $L[(\hat B^1)\cdot(\hat B^1)]+\cl A.$ because
$$\eqalignno{f_ih_{i-n}+h_if_{i-n}+k_i=&(f_{i-n}-f_i)(h_i-h_{i-n})
+(k_i+2f_ih_i)+(f_{i-n}h_{i-n}-f_ih_i)\cr}$$
and the last term is clearly exact. Thus $L[(\hat B^1)\cdot(\hat B^1)]
+\cl A.\subset\hat Z^1$ has nonempty intersection with each
cohomology class. We show that \chop
$\Big(L[(\hat B^1)\cdot(\hat B^1)]+\cl A.\Big)
\cap\hat B^1=\{0\}$. Choose an $\omega$ of the form (5.7) which
is exact: $\omega=\hat d\varphi$, for $\varphi=\{g_i\}_{i\in{\bf Z}}$,
so
$$\sum_\ell^L(f_{i-n}^\ell-f^\ell_i)
(h_{i-n}^\ell-h^\ell_i)+k_i=g_{i-n}-g_i\eqno{(5.9)}$$
for all $i,\,n\in{\bf Z}$. For $n=0$ we see $k_i=0$. Let $n\to\infty$
to find $\sum\limits_\ell^Lf_i^\ell h_i^\ell=-g_i$, so (5.9) simplifies
to
$$\sum_\ell^L(2f_{i-n}^\ell h^\ell_{i-n}-f_i^\ell h^\ell_{i-n}
-f^\ell_{i-n}h^\ell_i)=0\qquad\forall\,i,\,n\in{\bf Z}\,.$$
Replace $i$ with $i+n$ and take the limit $n\to\infty$ to find:\chop
$2\sum\limits_\ell^Lf_i^\ell h_i^\ell=0$, hence $g_i=0$,
so $\omega=0$.

To compute $\hat H^2(\cl A.)$, we start with the exact two--forms
(5.2\rn1). Since a general one--form $\varphi$ has the form
$$\varphi(\alpha^n)=\Big\{\sum_\ell^Lf_i^\ell h^\ell_{i-n}
+g_{i-n}+k_i\Big\}_{i\in{\bf Z}}$$
substitution of this into (5.2\rn1) produces
$$\eqalignno{(\hat d\varphi)(\alpha^n,\,\alpha^m)=&
\hlf\sum_\ell^L\Big\{(f_{i-m}^\ell
-f_i^\ell)(h_{i-n}^\ell-h_i^\ell)-(f_{i-n}^\ell-f_i^\ell)
(h^\ell_{i-m}-h_i^\ell)\Big\}_{i\in{\bf Z}}.\qquad(5.10)\cr}$$
Denote the space of these by $\hat B^2$.
Now a general two--form has an expression
$$\eqalignno{\omega(\alpha^n,\,\alpha^m)&=\sum_\ell^L\Big\{
f_i^\ell(h_{i-n}^\ell g_{i-m}^\ell-h_{i-m}^\ell g_{i-n}^\ell)
+(t^\ell_{i-n}-t^\ell_{i-m})\cr
&\qquad+(u_{i-n}^\ell v_{i-m}^\ell-v_{i-n}^\ell u_{i-m}^\ell)
+r_i^\ell(s^\ell_{i-n}-s^\ell_{i-m})\Big\}_{i\in{\bf Z}}\cr}$$
where $f^\ell,\,h^\ell,\,g^\ell,\,u^\ell,\,v^\ell,\,r^\ell,\,
s^\ell,\,t^\ell\in\cl A.$. Observe that by regrouping we can
readjust the r--s part to make the u--v part exact, and that the
r--s and t--parts are already closed by (5.3). Thus in the
closure equation $0=(\hat d\omega)(\alpha^n,\,\alpha^m,\,\alpha^k)$
we only need to consider the f--h--g part.
So now the closure equation for $\omega$ reads:
$$\eqalignno{&0=(\hat d\omega)(\alpha^n,\,\alpha^m,\,\alpha^k)\cr
&=\sum_\ell^L\Big\{f_i^\ell(h_{i-k}^\ell g_{i-m}^\ell-
h_{i-m}^\ell g^\ell_{i-k}+h^\ell_{i-n}g^\ell_{i-k}-h^\ell_{i-k}
g^\ell_{i-n}
+h^\ell_{i-m}g^\ell_{i-n}-h^\ell_{i-n}g^\ell_{i-m})\cr
&\qquad+f^\ell_{i-n}(h^\ell_{i-m}g^\ell_{i-k}-h^\ell_{i-k}g^\ell_{i-m}
+h^\ell_{i-k}g^\ell_i-h^\ell_i g^\ell_{i-k}
+h^\ell_i g^\ell_{i-m}-h^\ell_{i-m}g^\ell_i)\cr
&\qquad+f_{i-m}^\ell(h^\ell_{i-k}g^\ell_{i-n}-h^\ell_{i-n}g^\ell_{i-k}
+h^\ell_i g^\ell_{i-k}-h^\ell_{i-k}g^\ell_i
+h^\ell_{i-n}g^\ell_i-h^\ell_i g^\ell_{i-n})\cr
&\qquad+f^\ell_{i-k}(h^\ell_{i-n}g^\ell_{i-m}-h^\ell_{i-m}g^\ell_{i-n}
+h^\ell_{i-m}g^\ell_i-h^\ell_i g^\ell_{i-m}
+h^\ell_ig^\ell_{i-n}-h^\ell_{i-n}g^\ell_i)\Big\}_{i\in{\bf Z}}\cr}$$
for all $n,\,m,\,k\in{\bf Z}$. Now let $k\to\infty$ and regroup to find:
$$\eqalignno{0=&\sum_\ell^L\Big\{f^\ell_i(h^\ell_{i-m}g^\ell_{i-n}
-h^\ell_{i-n}g^\ell_{i-m})+h^\ell_i(f^\ell_{i-n}g^\ell_{i-m}
-f^\ell_{i-m}g^\ell_{i-n})\cr
&\qquad\qquad\qquad+g^\ell_i(f^\ell_{i-m}h^\ell_{i-n}
-f^\ell_{i-n}h^\ell_{i-m})\Big\}_i\;\;\forall\,n,\,m\,.\cr}$$
More compactly, it says that for all $i,\,j,\,k\in{\bf Z}$:
$$\eqalignno{0&=\sum_\ell^L\Big\{f_i^\ell(h^\ell_jg^\ell_k-
h^\ell_kg^\ell_j)+h^\ell_i(f^\ell_kg^\ell_j-f^\ell_jg^\ell_k)
+g^\ell_i(f^\ell_jh^\ell_k-f^\ell_kh^\ell_j)\Big\}\cr
&=\sum_\ell^L\sum_{\sigma\in{\bf S}_3}\epsilon^\sigma f^\ell_{\sigma(i)}
h^\ell_{\sigma(j)}g^\ell_{\sigma(k)}\cr
&=P_-\,S_{ijk}\;,&(5.11)\cr}$$
where we think of ${\bf S}_3$ as permutations acting on the set
$\{i,\,j,\,k\}$, and we used the notation
$S_{ijk}:=\sum\limits_\ell^Lf_i^\ell h^\ell_jg^\ell_k$ and
$P_-=\sum\limits_{\sigma\in{\bf S}_3}\epsilon^\sigma$
which is idempotent.
 On comparing equation (5.11) with the closure
equation above for $\omega$, we see that (5.11) is also sufficient
for $\omega$ to be closed. (Note that if one of $f,\,h$ or $g$ is a
linear combination of the others for each $\ell$, e.g.
$f^\ell_i=\alpha^\ell h^\ell_i+\beta^\ell g^\ell_i$ for
each $i$ and $\ell$, then the closure equation will be
satisfied.) Thus $S_{ijk}\in(P_-)^\perp$, i.e. $S_{ijk}$
is symmetric with respect to some pair of indices in
$\{i,\,j,\,k\}$, so a closed form can always be written in
the form:
$$\eqalignno{\omega(\alpha^n,\,\alpha^m)&=\sum_\ell^L\Big\{
f^\ell_i(h^\ell_{i-n}g^\ell_{i-m}-h^\ell_{i-m}g^\ell_{i-n})
-h^\ell_i(g^\ell_{i-n}f^\ell_{i-m}-g^\ell_{i-m}h^\ell_{i-n})\cr
&\quad+(t^\ell_{i-n}-t^\ell_{i-m})+r^\ell_i(s^\ell_{i-n}
-s^\ell_{i-m})\Big\}_i+(\hat d\varphi)(\alpha^n,\,\alpha^m)\cr}$$
where the added exact form takes care of the u--v part in the
original expression. Conversely, such a form is always closed.
To now examine the factor space $\hat H^2$, we want to write
an $\omega$ in terms of products of exact one--forms
(having (5.7) in mind). A small calculation shows that by
absorbing the cross--terms into the r--s and t--parts, we can write
any closed two--form, up to an exact form as:
$$\eqalignno{\omega(\alpha^n,\,\alpha^m)&=\sum_\ell^L\Big\{
f^\ell_i\big[(h^\ell_{i-n}-h^\ell_i)(g^\ell_{i-m}-g^\ell_i)
-(g^\ell_{i-n}-g^\ell_i)(h^\ell_{i-m}-h^\ell_i)\big]\cr
&\qquad-h^\ell_i\big[(g^\ell_{i-n}-g^\ell_i)(f^\ell_{i-m}-f^\ell_i)
-(f^\ell_{i-n}-f^\ell_i)(g^\ell_{i-m}-g^\ell_i)\big]\cr
&\qquad\quad+(t^\ell_{i-n}-t^\ell_{i-m})+r^\ell_i(s^\ell_{i-n}
-s^\ell_{i-m})\Big\}_{i\in{\bf Z}}\qquad (5.12)\cr}$$
Notice that each of the square brackets is an exact form.
Denote the space of two--forms having an expression as in (5.12)
by $\cl Q.$. Then we show there is a linear bijection between
$\cl Q.$ and $\hat H^2$, which is done by proving
$\cl Q.\cap\hat B^2=\{0\}$.
Let $\omega\in\cl Q.$, so it has an expression (5.12), and observe that
due to the coefficients $f^\ell_i,\,h^\ell_i,\,r^\ell_i,$ we have
$$\lim_{k\to\infty}\alpha^{-k}\omega(\alpha^{n+k},\,\alpha^{m+k})
=\sum_\ell^L(t^\ell_{i-n}-t^\ell_{i-m})\quad\forall\,n,\,m\in{\bf Z}\,.$$
However, for an exact form as in (5.10), say
$$(\hat d\varphi)(\alpha^n,\,\alpha^m)=\sum_\ell^L\Big\{(p^\ell_{i-m}
-p^\ell_i)(q^\ell_{i-n}-q^\ell_i)-(q^\ell_{i-m}-q^\ell_i)(p^\ell_{i-n}
-p^\ell_i)\Big\}_i\;\;,$$
we have that
$$\eqalignno{\lim_{k\to\infty}\alpha^{-k}(\hat d\varphi)(\alpha^{n+k},
\alpha^{m+k})&=\lim_{k\to\infty}\sum_\ell^L\Big\{(p^\ell_{i-m}-p^\ell_{i+k})
(q^\ell_{i-n}-q^\ell_{i+k})\cr
&\quad\qquad\qquad-(q^\ell_{i-m}-q^\ell_{i+k})(p^\ell_{i-n}
-p^\ell_{i+k})\Big\}_i\cr
&=\sum_\ell^L\big\{p^\ell_{i-m}q^\ell_{i-n}-q^\ell_{i-m}p^\ell_{i-n}
\big\}_i&(5.13)\cr}$$
and this clearly goes to zero when either $n$ or $m\to\infty$,
whereas $\lim\limits_{m\to\infty}{(t^\ell_{i-n}-t^\ell_{i-m})}
=t^\ell_{i-n}$. Thus if $\omega$ is exact, $t^\ell_i=0$ for all
$\ell$ and $i$, and so $\lim\limits_{k\to\infty}\alpha^{-k}\omega(
\alpha^{n+k},\,\alpha^{m+k})=0$, and (5.13) can only be zero
when $\hat d\varphi=0$. So we have proven:\chop
{\bf Claim 5.14:} There is a linear bijection between $\cl Q.$
and $\hat H^2(\cl A.)$.\chop

\beginsection 6. Examples: (II) A Lie Group Action on a Noncommutative
Algebra.

Next we wish to do a simple noncommutative example, but since
the exact one--forms $\hat dA$ played such an important role in
(5.7) and (5.12), want to exploit these explicitly.
Recall that in differential geometry an n--form has an expression
$$\omega=\sum_if_i\,dg_i^1\wedge\cdots\wedge dg_i^n$$
and in Connes' differential envelope over an algebra $\cl A.$, an
n--form is a linear combination of formal expressions
$$\omega=(a_0+\lambda 1)\,da_1\,da_2\ldots da_n$$
where $a_i\in\cl A.,\;\lambda\in{\bf C}$, and given any
 monomial made up from
$a_i\in\cl A.$ and $db_i$, $b_i\in\cl A.$, we can convert it
to this form using the assumption that $d$ is a graded derivation on
the differential envelope.
In the present integrated differential geometry, we wish to
get as close as possible to such an expression of a general
n--form. Recall that a zero--form $\varphi$ is just an element
$A\in\cl A.$. Then $(\hat d\varphi)(\alpha)=\alpha(A)-A=:\hat d_\alpha
A$ for $\alpha\in G$. Now
$$\hat d_\alpha(AB)=\alpha(A)\,\alpha(B)-AB
=\hat d_\alpha(A)\cdot\hat d_\alpha(B)+\hat d_\alpha(A)\cdot B
+A\,\hat d_\alpha(B)$$
which can be thought of as an integrated form of the Leibniz rule.
This rule has already been used in Cuntz's algebra of formal differences
[3].
So by
$$\hat d_\alpha(A)\cdot B=\hat d_\alpha(AB)-\hat d_\alpha(A)\cdot
\hat d_\alpha(B)-A\,\hat d_\alpha(B)\;,\eqno{(6.0)}$$
we can convert any expression of the form
$$\omega(\alpha)=\sum_iA_i^0\prod_{j=1}^{L_i}\hat d_\alpha(B_i^j)
A_i^j\eqno{(6.1)}$$
to the form
$$\omega(\alpha)=\sum_iC_i^0\prod_{j=1}^{K_i}\hat d_\alpha(E_i^j)\;.
\eqno{(6.2)}$$
Since furthermore any general one--form
$$\omega(\alpha)=\sum_iA_i^0\prod_{j=1}^{L_i}\alpha(B_i^j)A_i^j$$
(cf. (3.1)) can be written in the form (6.1), we conclude that
every one--form has an expression (6.2), which comes close to
the expression of a one--form for differential geometry.
For n--forms $\omega\in\hat\Omega^n$ we have likewise that they
can be expressed in the form:
$$\omega(\alpha_1,\ldots,\,\alpha_n)=\sum_{\sigma\in{\bf S}_n}
\epsilon^\sigma\sum_iA^0_i\prod_{j=1}^{L_i}(\hat dB^j_i)(
\alpha\s\sigma(s_i(j)).)$$
where $s_i:\{1,\,2,\ldots,\,L_i\}\to\{1,\,2,\ldots,\,n\}$
is a map and $L_i\geq n$.

Exact one--forms are of the type $\omega(\alpha)=\alpha(A)-A$,
$A\in\cl A.$, so if we use the expression for a one--form
$$\omega(\alpha)=\sum_{n=0}^N\sum_{i=0}^{K_n}A_i^n
\prod_{j=1}^n\hat d_\alpha(B_i^j)\eqno{(6.3)}$$
then the closure equation (5.1\rn2) is
$$\eqalignno{0&=\sum_n\sum_{i=0}^{K_n}\Big\{\alpha'(A_i^n)
\prod_{j=1}^n\big(\alpha(B_i^j)-\alpha'(B_i^j)\big)
-\alpha(A_i^n)\prod_{j=1}^n\big(\alpha'(B_i^j)-B_i^j\big)\cr
&\qquad\quad-\alpha'(A_i^n)\prod_{j=1}^n\big(B_i^j-\alpha'(B_i^j)\big)
+\alpha(A_i^n)\prod_{j=1}^n\big(B_i^j-\alpha(B_i^j)\big)\cr
&\qquad\qquad+A_i^n\prod_{j=1}^n\big(\alpha'(B_i^j)-B_i^j\big)
-A_i^n\prod_{j=1}^n\big(\alpha(B_i^j)-B_i^j\big)\Big\}\cr
&=\sum_n\sum_{i=0}^{K_n}\Big\{\big(\alpha'(A_i^n)-A_i^n\big)\prod^n_{j=1}
\big(\alpha(B_i^j)-B_i^j\big)+
\big(A_i^n-\alpha(A_i^n)\big)\prod_{j=1}^n\big(\alpha'(B_i^j)-B_i^j\big)
\Big\}\cr
&=\sum_n\sum_{i=0}^{K_n}\Big\{\hat d_{\alpha'}(A_i^n)\prod_{j=1}^n
\hat d_\alpha(B_i^j)-\hat d_\alpha(A_i^n)\prod_{j=1}^n\hat d_{\alpha'}
(B_i^j)\Big\}&(6.4)\cr}$$
for all $\alpha,\,\alpha'\in G$.
This equation for closure of one--forms has an interesting
resemblance to the closure condition for ordinary one--forms
in differential geometry. Next consider two--forms
which we know by (5.2) to be exact when there is a $\varphi\in
\hat\Omega^1$ such that
$$\omega(\alpha_1,\,\alpha_2)=\alpha_2\big(\varphi(\alpha_2^{-1}\cdot
\alpha_1)-\varphi(\alpha_2^{-1})\big)-\alpha_1\big(\varphi(\alpha_1^{-1}
\cdot\alpha_2)-\varphi(\alpha_1^{-1})\big)+\varphi(\alpha_2)-
\varphi(\alpha_1)$$
so on substituting in the canonical form:
$$\varphi(\alpha)=\sum_n\sum_{i=0}^{K_n}E_i^n\prod_{j=1}^n
\hat d_\alpha(D_i^j)\;,$$
we find that for all $\alpha_i\in G$:
$$\omega(\alpha_1,\,\alpha_2)=\sum_n\sum_{i=0}^{K_n}\Big\{
\hat d\s\alpha_2.(E_i^n)\prod_{j=1}^n\hat d\s\alpha_1.(D_i^j)
-\hat d\s\alpha_1.(E_i^n)\prod_{j=1}^n\hat d\s\alpha_2.
(D_i^j)\Big\}\quad(6.5)$$
The closure equation for two--forms in canonical form is very
messy, and we omit it.

We are now ready to attempt to find $\hat H^1(\cl A.)$ in a simple
noncommutative case.
Let $\cl A.=M_2({\bf C})$, and $G=\aut_0\cl A.\cong (U(2)/{\bf T})_0$,
the connected component of the identity of the
 automorphism group. Because $G$ is a
Lie group, we will be able to use differentiation at zero on
the one--parameter groups, to obtain a map $\tau$ from $\hat H^n(\cl A.)$
to infinitesimal cohomology for $\cl A.$. The action of $G$ on $\cl A.$
in terms of one--parameter groups is
$$\alpha_t(A)=e^{iBt}Ae^{-iBt}\qquad\forall\,t\in{\bf R},\;A\in\cl A.,\;
B=B^*\in\cl A.\;.$$
Now for a selfadjoint matrix $B$ we have either
$$B=\left(\matrix{\gamma&0\cr 0&\beta\cr}\right)\quad\hbox{or}\quad
B=r\left(\matrix{\gamma&e^{i\theta}\cr e^{-i\theta}&\beta\cr}\right)$$
where $\gamma,\,\beta,\,r,\,\theta\in\r$. Since $r$ can be absorbed
into the $t$, we will set $r=1$ henceforth.
So for {\bf Case 1} where $B=\left(\matrix{\gamma&0\cr 0&\beta\cr}\right)$,
we have obviously
$$U_t=\exp itB=\left(\matrix{e^{it\gamma}&0\cr 0&e^{it\beta}\cr}\right)\;.
\eqno{(6.6)}$$
\def\M#1,#2.#3,#4.{\left(\matrix{#1&#2\cr #3&#4\cr}\right)}
Now $B=\M\gamma,\exp i\theta.\exp(-i\theta),\beta.$ has eigenvalues
$E_1=\lambda+\mu$, $E_2=\chop\lambda-\mu$ where $\lambda:=\hlf(\gamma+\beta)
\in\r$, $\mu:=\hlf\sqrt{(\gamma-\beta)^2+4}\in[1,\,+\infty)$. So on
exponentiation we find for {\bf Case 2} where $\gamma\geq\beta$ that
$$\eqalignno{
&U_t=e^{itB}={ie^{it\lambda}\over\mu}\M\sqrt{\mu^2-1}\sin t\mu
-i\mu\cos t\mu,e^{i\theta}\sin t\mu.e^{-i\theta}\sin t\mu,
-\sqrt{\mu^2-1}\sin t\mu-i\mu\cos t\mu.\qquad(6.7)\cr}$$
and for
{\bf Case 3} where $\gamma<\beta$ we have
$$\eqalignno{U_t=e^{itB}={ie^{it\lambda}\over\mu}\M-\sqrt{\mu^2-1}\sin t\mu
-i\mu\cos t\mu,e^{i\theta}\sin t\mu.e^{-i\theta}\sin t\mu,
\sqrt{\mu^2-1}\sin t\mu-i\mu\cos t\mu.\qquad(6.8)\cr}$$
Thus $(\hat dA)(\alpha_t)=U_tAU_{-t}-A=\exp(it\,{\rm ad}\,B)(A)-A$,
and hence for an $A=\M a,b.c,d.$ we get in case 1:
$$(\hat dA)(\alpha_t)=\M 0,b(e^{it(\gamma-\beta)}-1).
c(e^{it(\beta-\gamma)}-1),0.\eqno{(6.9)}$$
and when $\alpha_t$ is case 2 we have that $(\hat dA)(\alpha_t)$ is
$$\eqalignno{&\sin t\mu\M \scriptstyle{
q\sin t\mu+iy\cos t\mu\qquad\Big\|\qquad
e^{i\theta}[-(\mu y+q\sqrt{\mu^2-1})\sin t\mu +i(\mu q+y
\sqrt{\mu^2-1})\cos t\mu]},.\scriptstyle{e^{-i\theta}[(\mu y-q\sqrt{\mu^2-1})
\sin t\mu-i(\mu q-y\sqrt{\mu^2-1})\cos t\mu]\qquad\Big\|\qquad
-q\sin t\mu+iy\cos t\mu},. \quad(6.10)\cr}$$
where $y=\mu^{-2}(be^{-i\theta}-ce^{i\theta})$ and
$q=\mu^{-2}[d-a+(e^{-i\theta}b+e^{i\theta}c)\sqrt{\mu^2-1}]$.
When $\alpha_t$ is case 3 we find that $(\hat dA)(\alpha_t)$ is
$$\eqalignno{&
\sin t\mu\M\scriptstyle{
 q'\sin t\mu+iy\cos t\mu\qquad\Big\|\qquad
e^{i\theta}\big[(-\mu y+q'\sqrt{\mu^2-1})\sin t\mu+
i(\mu q'-y\sqrt{\mu^2-1})\cos t\mu\big]},. \scriptstyle{
e^{-i\theta}\big[(\mu y+q'\sqrt{\mu^2-1})\sin t\mu-i(\mu q'+
y\sqrt{\mu^2-1})\cos t\mu\big]\qquad\Big\|\qquad
-q'\sin t\mu+iy\cos t\mu},.\quad(6.11)\cr}$$
where $y$ is as above, and $q':=\mu^{-2}[d-a-(e^{-i\theta}b
+e^{i\theta}c)\sqrt{\mu^2-1}]$.
Now observe that the power series of $(\hat dA)(\alpha_t)$
in powers of $t$ has lowest order one.
On substitution of an $\omega$ of the form (6.3) into the closure
relation (6.4), we find that for all $\wt\alpha_t,\;\alpha_s\in G$
we have
$$\eqalignno{0&=(\hat d\omega)(\alpha_s,\,\wt\alpha_t)\cr
&=\sum_n\sum_{i=0}^{K_n}\Big\{(\hat dA_i^n)(\wt\alpha_t)
\prod_{j=1}^n(\hat dB_i^j)(\alpha_s)-(\hat dA_i^n)(\alpha_s)\prod_{
j=1}^n(\hat dB_i^j)(\wt\alpha_t)\Big\}&(6.12)\cr}$$
which can be expressed as a polynomial in $s$ and $t$ with constant matrix
coefficients, and we see for each order of $s^kt^\ell$ that the
coefficient must vanish. The coefficient of the
lowest order of (6.12) (using the fact that
each $(\hat dA)(\alpha_t)$ starts with order one in $t$) is found from
$$0=\lim_{t\to 0}{1\over t}
\sum_{i=0}^{K_0}(\hat dA_i^0)(\alpha_t)=\lim_{t\to 0}{1\over t}
\hat d\Big(\sum_{i=0}^{K_0}A_i^0\Big)(\alpha_t)$$
for all
$\alpha_t\in G$. On setting $A^0:=
\sum\limits_{i=0}^{K_0}A_i^0=\M a,b.c,d.$ we find from  (6.9)
that $b=c=y=0$ and $q=(d-a)/\mu^2$, so (6.10) becomes
for $\lim\limits_{t\to 0}\f 1,t.(\hat dA^0)(\alpha_t)$:
$$\eqalignno{&\lim_{t\to 0}{\sin t\mu\over t}
\M q\sin t\mu\qquad\Big\|
\qquad e^{i\theta}q[-\sqrt{\mu^2-1}\sin t\mu+i\mu\cos t\mu],.
-e^{-i\theta}q[\sqrt{\mu^2-1}\sin t\mu+i\mu\cos t\mu]\qquad
\Big\|\qquad-q\sin t\mu,.=0\cr}$$
so $q=0$, i.e. $d=a$, hence $A^0=aI$ and $\hat dA^0=0$.
 Similarly we obtain from
(6.11) the same conclusion. Now for the coefficient of the
$st\hbox{--term}$ in (6.12), using $A^0=aI$,
we find
$$0=\lim_{s\to 0}\lim_{t\to 0}{1\over st}
\sum_{i=0}^{K_1}\Big\{(\hat dA_i^1)(\wt\alpha_t)(\hat dB_i^1)(\alpha_s)
-(\hat dA_i^1)(\alpha_s)(\hat dB_i^j)(\wt\alpha_t)\Big\}
\eqno{(6.13)}$$
for all $\alpha_s$ and $\wt\alpha_t\in G$.
Now from (6.9) we see for $\alpha_s$ type 1:
$$\lim_{s\to 0}{1\over s}(\hat dA)(\alpha_s)=i\M 0,b(\gamma-\beta).
c(\beta-\gamma),0.$$
and from (6.8) for type 2:
$$\lim_{s\to 0}{1\over s}(\hat dA)(\alpha_s)=i\M y,e^{i\theta}
(\mu q+y\sqrt{\mu^2-1}).-e^{-i\theta}(\mu q-y\sqrt{\mu^2-1}),y.$$
whilst (6.11) produces for type 3:
$$\lim_{s\to 0}{1\over s}(\hat dA)(\alpha_s)=i\M y,e^{i\theta}
(\mu q'-y\sqrt{\mu^2-1}.-e^{-i\theta}(\mu q'+y\sqrt{\mu^2-1}),y.
\eqno{(6.14)}$$
So there are six possible substitutions to make into (6.13).
Observe that when both $\alpha_s,\,\wt\alpha_t$ are type 1,
(6.13) only produces an identity. For the rest, we collect the
results in the following claim, where we assume that
$$A_i^1=\M a_i,b_i.c_i,d_i.\qquad\hbox{and}\qquad
B_i^1=\M e_i,f_i.g_i,h_i.$$
\thrm Claim 6.15." With notation above we have that\chop
$(\rn1)$ if an $\omega$ of the form (6.3) is closed, then
$$\eqalignno{\sum_ig_ib_i&=\sum_ic_if_i&(6.15\rn1)\cr
\sum_ib_i(h_i-e_i)&=0=\sum_ig_i(d_i-a_i)&(6.15\rn2)\cr
\sum_ib_if_i&=0=\sum_ig_ic_i&(6.15\rn3)\cr
\sum_if_i(d_i-a_i)&=0=\sum_ic_i(h_i-e_i)&(6.15\rn4)\cr
0&=\sum_i(d_i-a_i)(h_i-e_i)&(6.15\rn5)\cr
\sum_ib_ig_i&=0=\sum_ic_if_i\;,&(6.15\rn6)\cr}$$
and there are no further conditions on $A_i^1$ and $B_i^1$.\chop
$(\rn2)$ if $\omega\in\hat\Omega^1$ is closed, its first
order term in $\hat d$, i.e. $\sum\limits_iA^1_i\hat d_\alpha(B_i^1)$,
is exact.\chop
$(\rn3)$ There is a linear bijection from $\hat H^1$ to the closed
forms of the type
$$\omega(\alpha)=aI+\sum_{n=2}^N\sum_{i=0}^{K_n}A_i^n\prod_{j=1}^n
\hat d_\alpha(B_i^j)\;.\eqno{(6.16)}$$"
Let $\wt\alpha_t$ be type 1 and $\alpha_s$ type 2, then we obtain for
the st--coefficient (6.13) that
$$\eqalignno{\sum_{i=0}^{K_1}&\M 0,b_i(\wt\gamma-\wt\beta).
c_i(\wt\beta-\wt\gamma),0.\M y_i^B\qquad\Big\|\qquad
e^{i\theta}(\mu q_i^B+y_i^B\sqrt{\mu^2-1}),.
-e^{-i\theta}(\mu q_i^B-y_i^B\sqrt{\mu^2-1})\qquad\Big\|\qquad y_i^B,.\cr
&=\sum_{i=0}^{K_1}\M y_i^B\qquad\Big\|\qquad
e^{i\theta}(\mu q_i^A+y_i^A\sqrt{\mu^2-1}),.
-e^{-i\theta}(\mu q_i^A-y_i^A\sqrt{\mu^2-1})\qquad\Big\|\qquad y_i^B,.
\M 0,f_i(\wt\gamma-\wt\beta).g_i(\wt\beta-\wt\gamma),0.\cr}$$
where
$$\eqalignno{q_i^A&:=\mu^{-2}[d_i-a_i+(e^{-i\theta}b_i+e^{i\theta}c_i)
\sqrt{\mu^2-1}]\cr
q_i^B&:=\mu^{-2}[h_i-e_i+(e^{-i\theta}f_i+e^{i\theta}g_i)
\sqrt{\mu^2-1}]\cr
y_i^A:=\mu^{-2}&(b_ie^{-i\theta}-c_ie^{i\theta})\;,\qquad\quad
y_i^B:=\mu^{-2}(f_ie^{-i\theta}-g_ie^{i\theta})\,.\cr}$$
Multiplying out and equating matrix entries we find:
$$\eqalignno{\sum_ib_i(\mu q_i^B-y_i^B\sqrt{\mu^2-1})&=
e^{i2\theta}\sum_ig_i(\mu q_i^A+y_i^A\sqrt{\mu^2-1})&(\rn1)\cr
\sum_iy_i^Bb_i&=\sum_iy_i^Af_i&(\rn2)\cr
\sum_ic_iy_i^B&=\sum_iy_i^Ag_i&(\rn3)\cr
\sum_if_i(\mu q_i^A-y_i^A\sqrt{\mu^2-1})&=
e^{i2\theta}\sum_ic_i(\mu q_i^B+y_i^B\sqrt{\mu^2-1})&(\rn4)\cr}$$
Expand $(\rn2)$ to find for all $\theta$:
$$\sum_i(f_ie^{-i\theta}-g_ie^{i\theta})b_i=\sum_i(b_ie^{-i\theta}
-c_ie^{i\theta})f_i$$
so on cancelling we obtain (6.15\rn1), which would also have
followed from $(\rn3)$. Next expand $(\rn1)$ and cancel to find
$$\eqalignno{\sum_ib_i\left(h_i-e_i+e^{-i\theta}f_i(1-\mu^{-1})\sqrt{\mu^2-1}
\right)&=e^{i2\theta}\sum_ig_i\left(d_i-a_i+e^{i\theta}c_i(1-
\mu^{-1})\sqrt{\mu^2-1}\right)\quad(*)\cr}$$
for all $\mu$ and $\theta$. In the case when $\mu=1$ we obtain for all
$\theta$ that $\sum\limits_ib_i(h_i-e_i)=e^{i2\theta}\sum\limits_i
g_i(d_i-a_i)$ from which we deduce (6.15\rn2). When we substitute
this back into $(*)$ when $\mu\not=1$ and cancel, we find
$\sum\limits_ib_if_ie^{-i\theta}=\sum\limits_ig_ic_ie^{i3\theta}$
for all $\theta$, from which we deduce (6.15\rn3). Similarly by
expanding $(\rn4)$ we obtain (6.15\rn4).\chop
Next, we consider the case where $\wt\alpha_t$ is type 1 and
$\alpha_s$ is type 3 in (6.13). Note first from (6.14) that type 2
is converted to type 3 by the substitutions $\theta\to\theta+\pi$
and $q\to -q'$. On application of these to $(\rn1-\rn4)$ we find
$$\eqalignno{\sum_i(-\mu q_i'^B-y_i^B\sqrt{\mu^2-1})&=e^{i2\theta}
\sum_ig_i(-\mu q_i'^A+y_i^A\sqrt{\mu^2-1})&(\rn1')\cr
\sum_iy_i^Bb_i&=\sum_iy_i^Af_i&(\rn2)\cr
\sum_ic_iy_i^B&=\sum_iy_i^Ag_i&(\rn3)\cr
\sum_if_i(-\mu q_i'^A-y_i^A\sqrt{\mu^2-1})&=e^{i2\theta}
\sum_ic_i(-\mu q_i'^B+y_i^B\sqrt{\mu^2-1})&(\rn4')\cr
\hbox{where:}\qquad\qquad q_i'^A&:=\mu^{-2}[d_i-a_i-
(e^{-i\theta}b_i+e^{i\theta}c_i)\sqrt{\mu^2-1}]\cr
q_i'^B&:=\mu^{-2}[h_i-e_i-(e^{-i\theta}f_i+e^{i\theta}g_i)
\sqrt{\mu^2-1}]\;.\cr}$$
On expanding $(\rn1')$ and $(\rn4')$ we find they are already
satisfied by virtue of (6.15\rn1--\rn4). Next we let both
$\alpha_s$ and $\wt\alpha_t$ be type 2 in (6.13):
$$\eqalignno{&\sum_i^{K_1}\M\wt y_i^A\qquad\Big\|\qquad
e^{i\wt\theta}(\wt\mu\wt q_i^A+\wt y_i^A\sqrt{\wt\mu^2-1}),.
-e^{-i\wt\theta}(\wt\mu\wt q_i^A-\wt y_i^A\sqrt{\wt\mu^2-1})
\qquad\Big\|\qquad\wt y_i^A,.
\vphantom{\Bigg|}
\M y_i^B\qquad\Big\|\qquad e^{i\theta}(\mu q_i^B+y_i^B\sqrt{\mu^2-1}),.
-e^{-i\theta}(\mu q_i^B-y_i^B\sqrt{\mu^2-1})\qquad\Big\|\qquad y_i^B,.\cr
\cr
=&\sum_i^{K_1}\M y_i^A\qquad\Big\|\qquad e^{i\theta}(\mu q_i^A+
y_i^A\sqrt{\mu^2-1}),.
-e^{-i\theta}(\mu q_i^A-y_i^A\sqrt{\mu^2-1})\qquad\Big\|\qquad y_i^A,.
\M\wt y_i^B\qquad\Big\|\qquad e^{i\wt\theta}(\wt\mu\wt q_i^B+
\wt y_i^B\sqrt{\wt\mu^2-1}),.
-e^{-i\wt\theta}(\wt\mu\wt q_i^B-\wt y_i^B\sqrt{\wt\mu^2-1})
\qquad\Big\|\qquad\wt y_i^B,.\quad(\oplus)\cr}$$
Multiplying out and equating matrix entries, we find for the
upper diagonal entry:
$$\eqalignno{\sum_i\Big[\wt y_i^Ay_i^B-&e^{i(\wt\theta-\theta)}
(\wt\mu\wt q_i^A+\wt y_i^A\sqrt{\wt\mu^2-1})(\mu q_i^B-y_i^B
\sqrt{\mu^2-1})\Big]\cr
&=\sum_i\Big[y_i^A\wt y_i^B-e^{i(\theta-\wt\theta)}
(\mu q_i^A+y_i^A\sqrt{\mu^2-1})(\wt\mu\wt q_i^B-\wt y_i^B
\sqrt{\wt\mu^2-1})\Big]\cr}$$
On expansion of this, and eliminating terms via equations (6.15\rn1--\rn4),
we get
$$\eqalignno{\sum_i\Big[b_ig_i(\mu-1)(\wt\mu-1)\sqrt{(\mu^2-1)(\wt\mu^2-1)}
\sin[2(\theta-\wt\theta)]&+(d_i-a_i)(h_i-e_i)\mu\wt\mu\sin(\theta-
\wt\theta)\Big]=0\cr}$$
for all $\mu,\,\wt\mu,\,\theta,\wt\theta$. On setting $\mu=\wt\mu=1$
we obtain (6.15\rn5), and on substituting it back and using (6.15\rn1)
we obtain (6.15\rn6). Now it is a straightforward verification
to check that the set of equations (6.15\rn1--\rn6) guarantee that
the matrix equation $(\oplus)$ is satisfied for all its entries,
and moreover for the two remaining choices $\alpha_s$ and $\wt\alpha_t$
being either both type 3 or one type 2 and the other type 3; we find
that the set of equations (6.15) are also sufficient for (6.13)
to hold. We omit the calculations.\chop
$(\rn2)$ From part $(\rn1)$ we know that a closed one--form has expression
$$\omega(\alpha)=aI+\sum_{i=0}^{K_1}A_i^1\hat d_\alpha(B_i^1)
+\sum_{n=2}^N\sum_{i=0}^{K_n}A_i^n\prod_{j=1}^n\hat d_\alpha(B_i^j)
\eqno{(6.17)}$$
where $A_i^1$ and $B_i^1$ satisfy equations (6.15).
Now observe from (6.10) and (6.11) that when $\alpha_t$ is either
type 2 or 3, the entries of $(\hat dB_i^1)(\alpha_t)$ consist of
linear combinations of $y^{B_i^1}=\mu^{-2}(f_ie^{-i\theta}-g_ie^{i\theta})$,
$q^{B_i^1}=\mu^{-2}(h_i-e_i+(f_ie^{-i\theta}+g_ie^{i\theta})\sqrt{\mu^2-1})$
and $q'^{B_i^1}=\mu^{-2}(h_i-e_i-(f_ie^{-i\theta}+g_ie^{i\theta})
\sqrt{\mu^2-1})$ and so in the expression $\sum\limits_{i=0}^{K_1}
A_i^1\hat d\s\alpha_t.(B_i^1)$ we see by eqs (6.15) that only
combinations involving $a_i$ and $d_i$ are nonzero, i.e.
$\sum\limits_iA_i^1\hat d\s\alpha_t.(B_i^1)=\sum\limits_i\M a_i,0.0,d_i.
\hat d\s\alpha_t.(B_i^1)$. In fact, using 6.15\rn2, \rn4 and \rn5
we have $a_i=d_i$, so $\sum\limits_iA_i^1\hat d_\alpha(B_i^1)=
\sum\limits_ia_i\hat d_\alpha(B_i^1)=\hat d_\alpha\Big(
\sum\limits_ia_iB_i^1\Big)$. When $\alpha_t$ is type 1,
we have from (6.9), by (6.15\rn6) that
$$\eqalignno{
\sum_i\M a_i,b_i.c_i,d_i.\M 0\qquad f_i(e^{it(\gamma-\beta)}-1)
\vphantom{\Big|},.
g_i(e^{it(\beta-\gamma)}-1)\qquad 0\vphantom{\Big|},.&
=\sum_i\M  0\qquad\quad\vphantom{\Big|}a_if_i(e^{it(\gamma-\beta)}-1),.
d_ig_i(\vphantom{\Big|}e^{it(\beta-\gamma)}-1)\qquad\quad 0,.\cr}$$
and so by (6.15\rn4) in this case too, we have that
$\sum\limits_iA_i^1\hat d_\alpha(B_i^1)=\hat d_\alpha(\sum\limits_i
a_iB_i^1)$, which is therefore true for all $\alpha\in G$.
Thus the first order term in $\hat d$ for a closed one--form
$\omega$ is exact.\chop
$(\rn3)$ We already know that by the preceding parts of the claim,
there is a closed form of the type (6.16):
$$\omega(\alpha)=aI+\sum_{n=2}^N\sum_{i=0}^{K_n}A_i^n
\prod_{j=1}^n\hat d_\alpha(B_i^j)$$
in each cohomology class. We only need to show that such a
closed one--form is exact iff it is zero. Observe from (6.9),
(6.10) and (6.11) that for all types of $\alpha_t$, the power
series of $\omega(\alpha_t)$ in $t$ has no first order term.
Now any exact one--form $\hat d(A)(\alpha_t)$ must necessarily
have a nonzero first order term, since otherwise we see from
the expression
$$\hat d(A)(\alpha_t)=\exp({\rm ad}\,itB)(A)-A$$
that its higher order terms are also zero. Thus a one--form
of the type (6.16) is exact iff it is zero.

\itemitem{\bf Remarks:}(1) So by this claim we can visualize
$\hat H^1$ as the linear space of closed one--forms of the type
(6.16). We will not here work out the conditions on the entries
of $A_i^n$ and $B_i^j$, $n\geq 2$ to ensure that such an
$\omega$ is closed.
\itemitem{(2)} Clearly for a closed $\omega$ of type
(6.16), we have ${d\over dt}\omega(\alpha_t)\Big|_{t=0}=0$,
hence the {\it infinitesimal} first cohomology for $G$ acting
on $\cl A.=M_2({\bf C})$ is zero.

\beginsection 7. Further Developments.

First some comments on relating integrated de Rham cohomology
to existing cohomologies for algebras and groups.\chop
$(\rn1)$ Hochschild cohomology for algebras is constructed from
cochains consisting of n--linear maps from an algebra $\cl A.$
to an $\cl A.\hbox{--module}$ $X$. This is quite different from
the cochains of integrated de Rham cohomology, consisting of
maps $\omega:G^n\to\cl A.$ where $G$ acts on $\cl A.$, so there
seems to be little connection. Moreover, Hochschild cohomology is intrinsic to
algebras, regardless of any group actions.\chop
$(\rn2)$ Group cohomology starts from cochains which are maps
$\varphi:G^n\to Y$ where $G$ is a group and $Y$ is a coefficient group
on which G acts. So given an action $G\subset{\rm Aut}\,\cl A.$ on an algebra,
we can regard $\cl A.$ with its additive structure as such a coefficient group.
In this case we can regard the cochains of integrated de Rham cohomology
$\omega:G^n\to\cl A.$  as a subset of the cochains of group cohomology with
coefficient group $\cl A.$. However which particular subset
it will be, depends on the algebraic structure of $\cl A.$.
Moreover, the group coboundary operator is relatively insensitive to
the action of $G$ on $\cl A.$, whilst $\hat d$ is extremely sensitive
to the action. So again, there seems to be little connection.\chop
$(\rn3)$ In Connes' differential envelope over an algebra $\cl A.$,
there is no reference to a group action. One may try to take care of this,
using tensor products and homomorphisms, but this is unlikely to succeed
for the following reasons:
\item{a)} in Connes' differential envelope, formal expressions of the
type:
$$\omega=(a_0+\lambda 1)\,da_1\,da_2\ldots da_n\eqno{(*)}$$
are the basic objects, whilst in our case, the maps $\omega:G^n\to\cl A.$
are basic, and the same map may have different expressions
$$\omega(\alpha_1,\ldots,\alpha_n)=(a_0+\lambda 1)\hat d\s\alpha_i.
a_1\ldots\hat d\s\alpha_k.a_m\eqno{(+)}$$
which we identify.
\item{b)} The expression $(*)$ in the differential envelope for an
n--form has precisely n factors $da_i$, whilst in our case, in
$(+)$ there is no upper bound on the (finite) number of
$\hat d\s\alpha_i.a$ factors which can occur for an n--form.
\item{c)} $d$, as an operator on the differential envelope
satisfies the graded Leibniz rule, whereas $\hat d$ satisfies (6.0).

\noindent$(\rn4)$ Cuntz in [3] defines an algebra of formal differences
in which the basic objects do satisfy (6.0) and the algebra consist of
formal products of these. There is no reference to a group action,
and it also appears difficult to connect to
integrated differential geometry for reason (a) above.\chop

The rest of the machinery of differential geometry is quite easy to
define in integrated differential geometry, for instance it
has been done for push--forwards, pull--backs, Lie derivatives,
principal fibre bundles and connections on them. In each case,
the integrated object
is defined in such a way that under the $\tau$ map on $C^\infty(M)$
it reduces to the usual object. The main application for such
an extension of differential geometry would be to Hamiltonian mechanics
and classical gauge theories. Whilst we can easily imitate the formal
structure of Hamiltonian mechanics in integrated differential geometry,
what is really needed is a way of doing actual Hamiltonian mechanics
using only structures of integrated differential geometry
(without reference to the infinitesimal level, i.e. the map $\tau$).
That is, from the Hamiltonian function and symplectic form (integrated),
we should obtain the same time evolution groups on $C^\infty(M)$
by such an integrated method, as that obtained by the usual
Hamiltonian mechanics. Thus far, such a method has been eluding
the author, and so we leave the further development of integrated
differential geometry for a future project.

\beginsection 8. Discussion.

Above, we have shown that for a manifold $M$, there is a
larger ``integrated differential geometry'' structure defined
on the action of ${\rm Diff}(M)$ on $C^\infty(M)$ such that
when we differentiate at zero along the one--parameter groups
we obtain ordinary differential geometry. This structure
generalised readily to all group actions on associative
algebras, and provided a chain complex from which we
could define ``integrated de Rham cohomology,''
thus establishing a set of new invariants for group actions.
This was applied to two examples;- calculating $\hat H^1$
and $\hat H^2$ for the shift operator acting on an algebra
of sequences, and finding $\hat H^1$ for the algebra
$M_2({\bf C})$ under its automorphism group.

Of the many possible directions for developing this structure further,
we note a few;- first, examining topological questions when
$G$ and $\cl A.$ are endowed with topologies; second, how this
structure intertwines with the covariant representation theory of the
group action, and thirdly, do a more difficult example, e.g.
${\rm Diff}(S^1)$ acting on $C(S^1)$.
For comparison with Connes' approach in an example,
a good example would be the action of the permutation group
on the algebra of functions on a discrete set, cf. [7].
 Apart from this,
there is the development of integrated Hamiltonian mechanics and
gauge theory, as mentioned above.

\beginsection Acknowledgements.

The idea for this paper occurred to me whilst working on a joint
project with Mark Gotay and Angas Hurst on geometric quantisation.
It was developed purely for my own pleasure, and is intended
as a minor footnote to that large and deep theory [1,2] developed
by Alain Connes and his co--workers.
I am also grateful to Norman Wildberger for his incredulous but
patient ear over many coffees, to Keith Hannabuss for harboring
me in Oxford where I tried getting the examples straight,
to prof. Cuntz for his remarks and preprints, and to Prof. R. Coquereaux
for his hospitality at the CNRS, Luminy.

\beginsection Bibliography.

\item{[1]} Connes, A.: Non-commutative differential geometry,
Publ. Math. IHES {\bf 62}, 257--360 (1985)\chop
Connes, A: Non--commutative differential geometry, Academic Press (in press).
\item{[2]} Cuntz, J.: Representations of quantized differential forms
in \chop non--commutative geometry. Preprint, Heidelberg 1994.
\item{[3]} Cuntz, J.: A survey of some aspects of non--commutative
differential geometry, Jber. d. Dt. Mat.--Verein. {\bf 95}, 60--84 (1993)
\item{[4]} Bratteli, O.: Derivations, dissipations and group actions on
C*--algebras, Springer Lect. Notes Math. 1229 (1986)
\item{[5]} Dubois--Violette, M., Kerner, R., Madore, J.:
Noncommutative differential geometry of matrix algebras,
J. Math. Phys. {\bf 31}, 316--322 (1990)
\item{[6]} Blackadar, B., Cuntz, J.: Differential Banach algebra norms
and smooth subalgebras of C*--algebras, J. Op. Theory {\bf 26}, 255--282 (1991)
\item{[7]} Dimakis, A., M\"uller--Hoissen, F.: Discrete differential
calculus, graphs, topologies and gauge theory, J. Math. Phys. to appear.

\bye